\documentclass[]{sn-jnl}
\usepackage{graphicx}%
\usepackage{multirow}%
\usepackage{amsmath,amssymb,amsfonts}%
\usepackage{amsthm}%
\usepackage{mathrsfs}%
\usepackage[title]{appendix}%
\usepackage{xcolor}%
\usepackage{textcomp}%
\usepackage{manyfoot}%
\usepackage{booktabs}%
\usepackage{algorithm}%
\usepackage{algorithmicx}%

\def\arcsec{\hbox{$^{\prime\prime}$}}
\def\arcmin{\hbox{$^{\prime}$}}

\newcommand{\apj}{Astrophys. J.}   
\newcommand{\aap}{Astron. Astrophys.}   
\newcommand{\mnras}{Mon. Not. R. Astron. Soc.}   
\newcommand{\nat}{Nature} 

\title{H$\alpha$-X-ray Surface Brightness Correlation for Filaments in Cooling Flow Clusters}

\author*[1,2,3]{\fnm{Valeria} \sur{Olivares}}\email{valeria.olivares@usach.cl}
\author[4,5,6]{\fnm{Adrien} \sur{Picquenot}}
\author[7]{\fnm{Yuanyuan} \sur{Su}}
\author[8]{\fnm{Massimo} \sur{Gaspari}}
\author[9,10]{\fnm{Marie-Lou} \sur{Gendron-Marsolais}}
\author[11]{\fnm{Fiorella L.} \sur{Polles}}
\author[12]{\fnm{Paul} \sur{Nulsen}}
\affil*[1]{\small Departamento de Física, Universidad de Santiago de Chile, Av. Victor Jara 3659, Santiago, Chile}
\affil[2]{\small Center for Interdisciplinary Research in Astrophysics and Space Exploration (CIRAS), Santiago, Chile}
\affil[3]{\small Astrophysics Branch, NASA Ames Research Center, MS 245-6, Moffett Field, CA 94035, USA}
\affil[4]{\small Department of Astronomy, University of Maryland, College Park, MD 20742, USA}
\affil[5]{\small X-ray Astrophysics Laboratory NASA/GSFC, Greenbelt, MD 20771, USA}
\affil[6]{\small Center for Research and Exploration in Space Science and Technology, NASA/GSFC, Greenbelt, MD 20771, USA}
\affil[7]{\small Department of Physics and Astronomy, University of Kentucky, 505 Rose Street, Lexington, KY 40506, USA}
\affil[8]{\small Department of Physics, Informatics and Mathematics, University of Modena and Reggio Emilia, 41125 Modena, Italy}
\affil[9]{\small Département de physique, de génie physique et d’optique, Université Laval, Québec, Q1V 0A4, QC, Canada}
\affil[10]{\small Instituto de Astrofísica de Andalucía, IAA-CSIC, Apartado 3004, 18080 Granada, España}
\affil[11]{\small SOFIA Science Center, USRA, NASA Ames Research Center, M.S. N232-12, Moffett Field, CA 94035, USA}
\affil[12]{\small ICRAR, University of Western Australia, 35 Stirling Hwy, Crawley, WA 6009, Australia}

\abstract{Massive galaxies in cooling flow clusters display clear evidence of feedback from Active Galactic Nuclei (AGN). Joint X-ray and radio observations have shown that AGN radio jets push aside the surrounding hot gas and form cavities in the hot intracluster medium (ICM). These systems host complex, kiloparsec-scale, multiphase filamentary structures, from warm ionized (10,000 K) to cold molecular ($<$100 K). These striking clumpy filaments are believed to be a natural outcome of thermally unstable cooling from the hot ICM, likely triggered by feedback processes while contributing to feeding the AGN via Chaotic Cold Accretion (CCA). However, the detailed constraints on the formation mechanism of the filaments are still uncertain, and the connection between the different gas phases has to be fully unveiled. By leveraging a sample of seven X-ray bright cooling-flow clusters, we have discovered a tight positive correlation between the X-ray surface brightness and the H$\alpha$ surface brightness of the filaments over two orders of magnitude, as also found in stripped tails.}

\begin{document}
\maketitle

\begin{figure}
    \centering
   \includegraphics[width=0.85\textwidth]{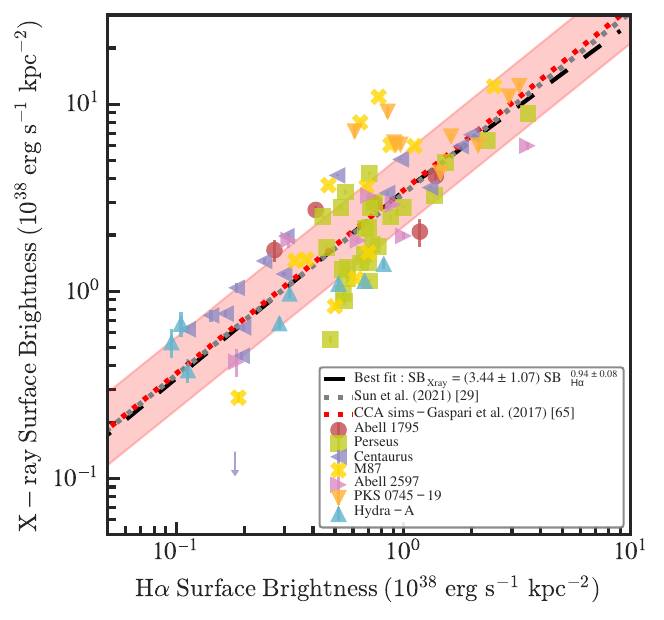}
\caption{ - H$\alpha$ -- X-ray surface brightness (SB) correlation for filaments in 7 strong cooling flow clusters. Each data point corresponds to a filament region for a given cooling flow cluster. X-ray surface brightness was computed within the 0.5\,--\,2.0 keV band using the X-ray image obtained using imaging decomposition method (pGMCA \cite{Picquenot19}). Errors and upper-limits correspond to 1$\sigma$. The dashed black line corresponds to the best fit to our dataset using Bayesian {\sc linmix} method \cite{Kelly07} with 1$\sigma$ errors in the slope and intercept. 
The dotted gray line shows the relation found for diffuse gas in stripped tails ($\rm SB_{Xray} = (3.33\pm0.34)~SB_{H\alpha}^{0.94\pm0.06}$, \cite{sun21}). The dotted red line is the relation predicted by CCA simulations, with the 1-$\sigma$ intrinsic scatter band superposed, which is consistent with our observational findings \cite{gaspari17}.} 
\label{fig:Xray_ha_ratios}
\end{figure}

Evidence of AGN feedback can be observed in cooling flow clusters, where powerful radio-emitting jets from the central galaxy create bubbles in the surrounding ICM \cite{mcnamara07,mcnamara00,fabian12,birzan20,birzan08,gitti10,Hlavacek-Larrondo15,olivares22b}. Multiphase filaments extended from the central galaxy may result from hot gas condensation triggered by AGN feedback \cite{mcdonald10,mcdonald12,conselice01,hamer16,salome03,salome11,edge01,russell19,vantyghem19,lim12,olivares19,olivares23,Ganguly23,Tamhane23,Vigneron24}. Spatial correlations between the X-ray and H$\alpha$ filaments in cooling-flow clusters have been observed since the 1990s \cite{sarazin92, sarazin92b, crawford05, sparks04, david17}. 
{However, those earlier observations did not establish a quantitative correlation between X-ray and H$\alpha$ surface brightness.}
Inspired by the universal X-ray to H$\alpha$ surface brightness {correlation} found in the stripped tails of jellyfish galaxies \cite{sun21}, we present a {quantitative} comparison of H-alpha and X-ray surface brightness in the filaments observed in multiple cooling flow clusters (Fig.~\ref{fig:Xray_ha_ratios}). These clusters were uniformly analyzed with a novel technique that allows the measurement of X-ray filament surface brightness over a large dynamic range.

We analyzed the deep \textit{Chandra} observations of 7 strong cooling flow clusters -- Perseus, M\,87, Centaurus, Abell\,2597, Abell\,1795, Hydra-A, and PKS\,0745-191 -- that display prominent multiphase filamentary structures (see Table~\ref{table:sample}). We first isolated the X-ray filamentary components from the underlying X-ray bright cool core and the complicated substructures at cluster centers such as X-ray cavities and sloshing cold fronts (see Section~\nameref{sec:method}). This was achieved by using a novel imaging decomposition method called the General Morphological Component Analysis (GMCA, \cite{Bobin15}), along with its updated versions pGMCA (Poisson Generalized Morphological Component Analysis \cite{Bobin20}) and \cite{Picquenot19} (which is designed to exploit X-ray data). The pGMCA method provides distinct X-ray components for each pixel, consisting of spatial and spectral information that make up the total X-ray emission of each cluster, including X-ray filaments, a diffuse X-ray halo that generally reveals a sloshing spiral, and X-ray cavities. Each of these components can be visually and spectrally identified (see Section~\nameref{sec:method}). All clusters have deep H$\alpha$ data (which traces the $\sim10,000$~K gas phase) from MUSE (Multi Unit Spectroscopic Explorer) or SITELLE (Spectromètre Imageur à Transformée de Fourier pour l'Etude en Long et en Large de raies d'Emission) integral field spectrograph (IFS) observations. 

\begin{figure}
    \centering
     \includegraphics[width=0.8\textwidth]{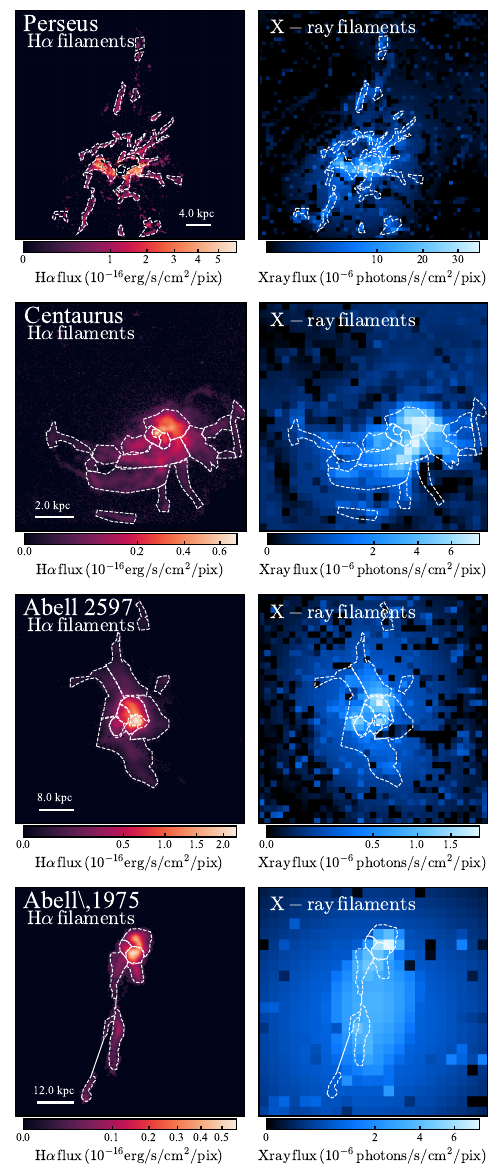}

\caption{ - {Distribution of the regions for each source}. H$\alpha$ (Left panel) and X-ray (Right panel) flux map of the filaments showing the different regions with white dashed lines used to measure the surface brightness of the H$\alpha$ and X-ray filaments. The central regions are not included in this study. Both panels show the same region.} \label{fig:regions}
\end{figure}

\begin{figure}
\centering

    \includegraphics[width=0.8\textwidth]{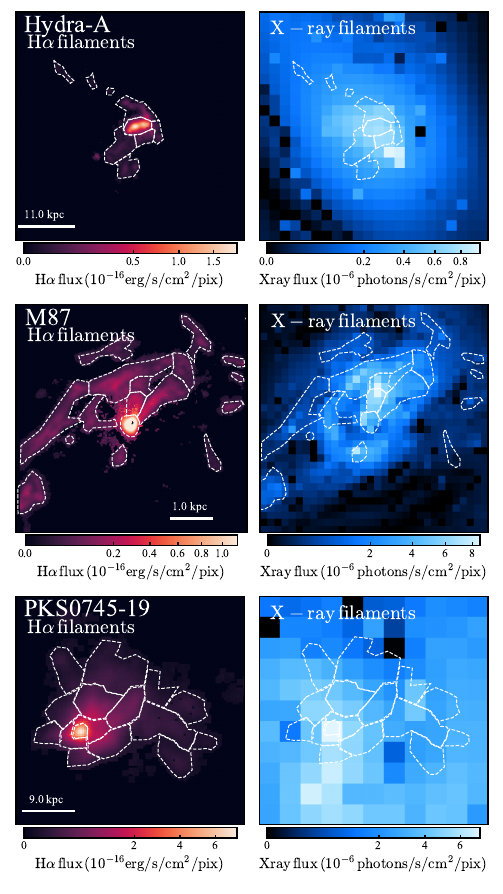}

\caption{  - \textbf{Distribution of the regions for each source}. H$\alpha$ (Left panel) and X-ray (Right panel) flux map of the filaments showing the different regions with white dashed lines used to measure the surface brightness of the H$\alpha$ and X-ray filaments. The central regions are not included in this study.} \label{fig:regions2}
\end{figure}

The filamentary structure for each cluster was divided into several regions (filaments) with sizes that fill a range from 0.1~kpc$^{2}$ to 27~kpc$^{2}$ (see \nameref{sec:supplementary}). 
The X-ray flux, surface brightness, and luminosity for each region have been measured in the 0.5--2.0~keV band, excluding the central region (2--4$\arcsec$), where the central bright AGN is located, and the point sources (see Section~\nameref{sec:method}, and Figures~\ref{fig:regions}, and \ref{fig:regions2}). The H$\alpha$ surface brightness and luminosity have been derived similarly, excluding stars and background sources. The surface brightness and luminosity of the X-ray filaments have been measured using the output image from the pGMCA method (see Section~\nameref{sec:method}). 

As shown in Figure~\ref{fig:Xray_ha_ratios}, the data unambiguously show a correlation (over 2 dex), between the surface brightness of the X-ray and H$\alpha$ emitting filaments in our sample, indicating a strong connection between the hot and warm gas phases. It is worth noting that the correlation vanishes when comparing the surface brightness of the H$\alpha$ filaments with the unfiltered X-ray image (X-ray halo), as the underlying ICM halo as well as the presence of cavities and sloshing spirals contribute to the total X-ray spectrum in cooling flow clusters, especially in the cluster core (\nameref{sec:supplementary}). 

The almost constant X-ray/H$\alpha$ surface brightness ratio of 4.1$\pm$2.4 along the different filaments of the entire sample (see Figure~\ref{fig:Xray_ha_ratios}) indicates that a local process must be responsible for the excitation of the gas, such as energetic particles like X-ray or cosmic rays \cite{ferland09}, shocks \cite{allen08}, turbulent mixing layers \cite{begelman90}, and reprocessing of the extreme ultraviolet (EUV) and X-ray radiation from the cooling plasma \cite{polles21}, while the distance to the central AGN does not play a significant role \cite{lim08,olivares19,hamer19}. Interestingly, this correlation agrees with the one found for the diffuse gas of stripped tails of Jellyfish galaxies traveling through the ICM \cite{sun21,poggianti19}, hinting similar processes may be at play. The intrinsic {random} scatter in our sample is larger, {between 0.04 and 0.11 in log10, or 10\% and 30\% (see \nameref{sec:supplementary})}, compared to the scatter in stripped bright tails \cite{sun21}, which is only 7--9\%. This difference could be due to residual emission from the X-ray halo and the non-detection of some X-ray filaments in some clusters (Section~\ref{sec:x-ray-fit}). However, it could also be due to some of the many physical processes that occur at the center of cooling flow clusters, such as variation of star formation rate that can boost H$\alpha$ emission, radio jets and cosmic rays. It {is likely} that some of these processes play a role in heating, ionizing, and shaping the properties of the H$\alpha$ and X-ray filaments.

We explore also whether the X-ray and H$\alpha$ filaments (hot and warm phases) are in pressure equilibrium (see \nameref{sec:supplementary}). For the optical emitting filaments of Centaurus and M87, we derived the electron density, $n_{\rm e}$, and temperature, $T_{\rm e}$, of the warm phase of each region. For that purpose, we used PyNeb \cite{pyneb} and the emission line ratios [SII]$\lambda$6716/[SII]$\lambda$6731, and [NII]$\lambda$5755 / ([NII]$\lambda$6548 + [NII]$\lambda$6583), respectively (see \nameref{sec:supplementary}). We obtained $n_{\rm e}$ values between 60\,--\,150~cm$^{-3}$ in the H$\alpha$ filaments. We derived an $T_{\rm e}$ of $\sim$10,000~K (with values between $\sim$9,000~K and $\sim$11,000~K) in the inner 5~kpc region of the nebulae of Centaurus, where the auroral [NII] emission line is detected. For M87, the auroral lines were detected in some filaments, we obtained $T_{\rm e}$ of $\sim$10,000~K, between 9,000~K and 12,000~K. For the filaments where the auroral [NII] emission line was undetected, we used an upper limit on the $T_{\rm e}=$10,000~K to derive the electron density.

Similarly, we compute the $n_{\rm e}$ and $T_{\rm e}$, by modeling of the X-ray spectrum of the X-ray filaments with PyXspec \cite{Arnaud96} (see Section~\nameref{sec:method}). We only perform the analysis of the filaments obtained with the pGMCA method for the cooling flow clusters with the deepest, high-resolution \textit{Chandra} observations. We found that the X-ray filaments in each cluster have a constant temperature but with a different value for each cluster, of 0.5~keV for Centaurus, 0.7~keV for M\,87, and 1.0~keV for Perseus (see Section~\ref{sec:x-ray-fit}). We note that previous studies, which utilized multitemperature fitting of the \textit{Chandra} observations, reveal comparable temperature values for the X-ray filaments (\cite{werner13,sanders07,sanders16}, see also Sec.~\nameref{sec:method} for a comparison between the two methods). The electron density of the X-ray filaments strongly depends on the assumed geometry and projection effects (see \nameref{sec:supplementary}). Assuming a cylindrical geometry, we find that the electron densities of the X-ray filaments lie between 0.02\,--\,0.45\,cm$^{-3}$, with a median value of 0.15~cm$^{-3}$. 

\begin{figure}
    \centering
    \includegraphics[width=0.95\textwidth]{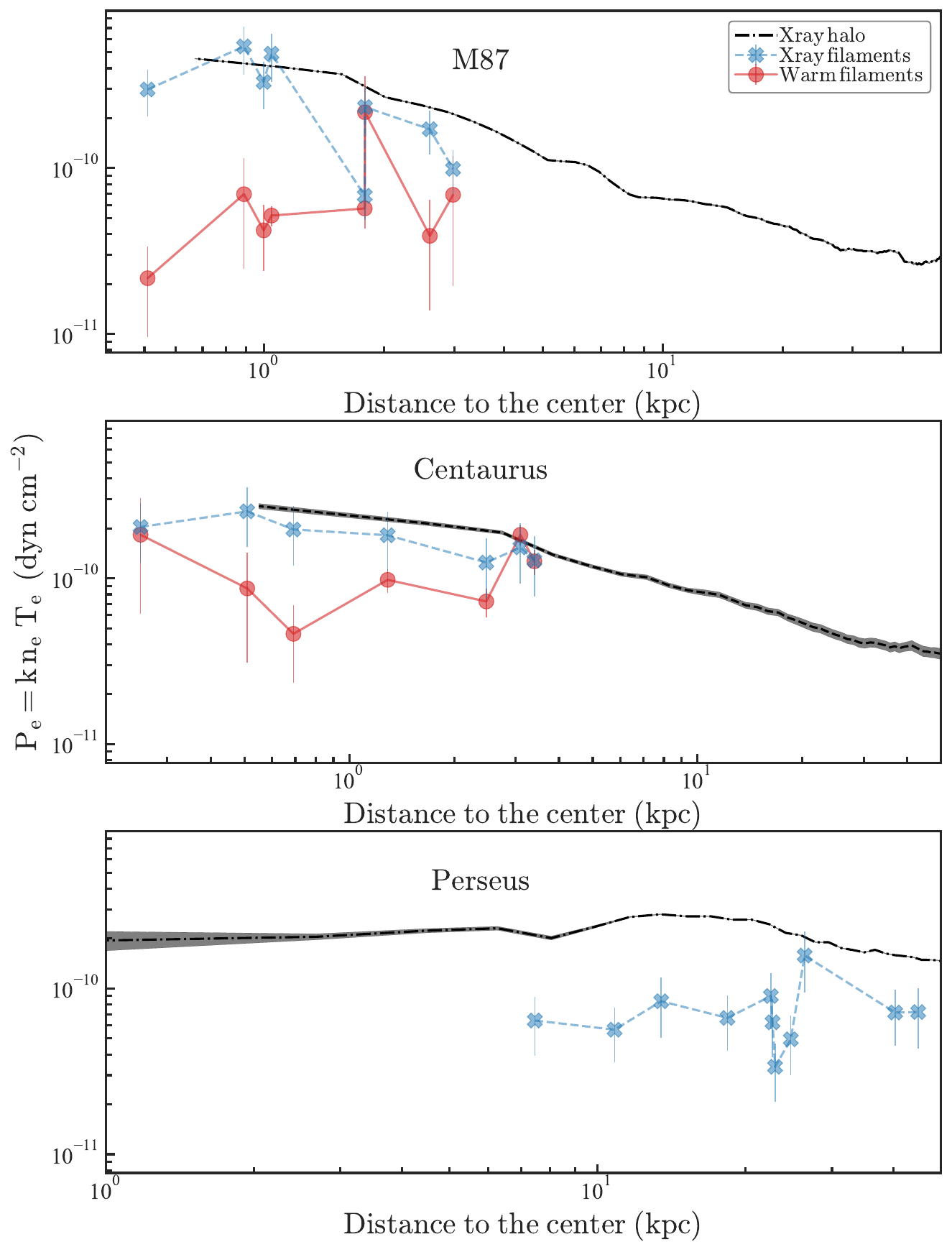}
    \caption{ - Electron pressure profiles of the different temperature gas phases of the filaments and X-ray halo for M87, Centaurus, and Perseus. The electron pressures are in {dyn~cm$^{-2}$}, with X-ray and H$\alpha$ filaments marked with crosses (blue) and circles (red), respectively. The electron pressure profiles of the X-ray halos are shown with a dashed black line for each source. Errors correspond to 1$\sigma$.}
    \label{fig:pressures}
\end{figure}

When comparing the electron pressure, $P_{\rm filaments} = k n_{\rm e} T_{\rm e}$, we find that the X-ray and H$\alpha$ filaments are out of pressure equilibrium (see Figure~\ref{fig:pressures}). The pressure of the X-ray filaments, $P_{\rm Xray, filaments}$, is almost a factor of a few, between 1 and 4, higher than the H$\alpha$ filaments, $P_{\rm H\alpha, filaments}$. As mentioned above, the electron pressure of the X-ray filaments is highly dependent on the geometry, and it could increase by a few, assuming the widths of filaments are smaller, $<$100~pc (see also \cite{sanders07,fabian08}). {Also, our pressure estimates for the X-ray filament could be underestimated if the cool X-ray gas does not cover the cylinder.} In contrast, projection effects may artificially increase the electron density and electron pressure of X-ray filaments. If a filament is placed along the line of sight, its volume will appear smaller than it actually is. This can cause us to overestimate its density and pressure by a factor of 1.2 if the filaments have an inclination angle of 45~degrees) with the line of sight (For more details, see \nameref{sec:supplementary}.)

The typical diffuse density of the X-ray halo is $n_{\rm e}\sim$0.1~cm$^{-3}$ within the region where the filaments are seen \cite{werner14}. As shown in Figure~\ref{fig:pressures}, neither the H$\alpha$ nor the X-ray filaments are in pressure equilibrium with the X-ray halo, since the pressure of the X-ray halo is almost always higher than that of the X-ray and H$\alpha$ filaments, $P_{\rm Xray, halo} \geq P_{\rm Xray, filament} > P_{\rm H\alpha, filament}$. Additionally, as seen in the Perseus cluster, the ratio in electron pressure between the X-ray halo and the X-ray filaments seems to increase by a factor of about 3, compared to Centaurus and M87, as we move to distances larger distances from the cluster center. {As shown in Figure~\ref{fig:pressures}, the electron pressure of the X-ray halo exceeds that of the H$\alpha$ filaments, indicating the presence of an additional non-thermal pressure component in the filaments. }

As implied by the narrow H$\alpha$ filaments in the Perseus cluster that have been observed with high-resolution Hubble Space Telescope (HST) \cite{fabian08}, likely physical mechanisms supporting the filaments against gravitational collapsing are magnetic fields. Indeed, most BCGs in cooling-flow clusters have low star formation activity (of a few 10s~M$_{\odot}$\,yr$^{-1}$, \cite{mittal15}), but an extensive reservoir of cold molecular gas \cite{salome03,edge01,salome08,olivares19,russell19}. Therefore, to avoid gravitational collapse of the clumpy filaments, a magnetic pressure, $P_{B}$ = $B^{2}/8$ $\pi$, between 1.6$\times$10$^{-11}$\,dyne\,cm$^{-2}$ and 1.4$\times$10$^{-10}$ dyne\,cm$^{-2}$, is needed to keep the different gas phases of the filaments in (total) pressure equilibrium, which corresponds to a magnetic field $B\sim$\,20\,--\,60~$\mu$G (\cite{fabian08}). 
The presence of a strong magnetic field (10--25~$\mu$G \cite{taylor06}) in filaments of cooling flow clusters has been previously inferred from geometry and widths of single extended filaments using high-resolution HST observations \cite{fabian08}, radio observations, and pressure balance arguments \cite{werner13}. At the same time, simulations predict a {higher magnetic field} in the filament than the X-ray halo, with a range of values depending on the phase of the filament, enabling significant pressure support for the filaments \cite{wang21, beckmann22, Fournier24}. In that context, the difference in pressure between the X-ray and optical filaments could also be explained by the difference in their magnetic field strength.

Turbulence energy could introduce another form of non-thermal pressure, $P_{\rm turb}\,=\,1/3\,\rho \sigma_{v}^{2}$, where $\sigma_{v}$ is the 3D turbulent velocity dispersion, and $\rho$ is the density of the hot gas. Given that velocity dispersions, $\sigma_{v}$, of 100~km~s$^{-1}$ are expected in the warm filaments, as found by line-width detections \cite{olivares19} based on MUSE observations, and supported by hydrodynamical simulations of CCA \cite{Gaspari_2018}, turbulent pressure represents another channel to prevent the condensing warm filaments from collapsing. {Assuming a typical electron density of 60~cm$^{-3}$ for the warm gas, a turbulent velocities roughly $\sigma_{v} \sim 100$ km~s$^{-1}$, and that the gas fully ionized, we can calculate the mass of the of ionized gas using the Eq.~13.8 from \cite{OsterbrockFerland2016}, and thus derived the a typical turbulent pressure of $P_{\rm turb} = 6.5\times10^{-12}$~dyne\,cm$^{-2}$. This additional non-thermal pressure can reduce the implied $B$ strength by 13~$\mu$G. 
Further, cosmic rays pressure could provide additional support, with some numerical simulations showing a significant cosmic ray pressure balancing the filament thermal component \cite{beckmann22,beckmann22b}.}

Our discovered correlation between the surface brightness of the X-ray and H$\alpha$ emitting filaments provides evidence for theoretical models of Chaotic Cold Accretion (CCA) and precipitation \cite{Gaspari_2013,sharma12,mccourt12,beckmann19, li14,li18,Voit_2017,voit19, Storchi-Bergmann_2019,wang21}.  In Figure \ref{fig:Xray_ha_ratios}, we show the quantitative prediction of CCA by leveraging high-resolution hydrodynamical simulations (\cite{gaspari17}; see \nameref{sec:supplementary}). The Bayesian fit retrieves a median slope and normalization consistent with our observations, with a slightly more elevated scatter, still within uncertainties.
This correlation arises naturally in CCA since nonlinear thermal instability (triggered via turbulence and AGN jets/bubbles \cite{mccourt12,sharma12,gaspari12,nobels22}) quickly induces the X-ray overdensities/filaments to condense down to the first stable H$\alpha$-emitting phase at $10^4$\,K, thus establishing tight spatial and thermo-kinematical correlations between the hot and cool gas \cite{Gaspari_2018}. Particularly relevant here is the tight correlation between density profiles with a logarithmic slope of -1 \cite{gaspari17}. While the H$\alpha$ phase has larger density normalization, this is counterbalanced by the lower volume filling and line emissivity.
Some of these condensed structures will be crucial to boost the accretion onto the central SMBH, stimulating an efficient self-regulated AGN feeding and feedback cycle \cite{gaspari20}. 
Conversely, any hot mode of accretion will show highly decorrelated phases (including brightness), since the multiphase gas is not causally connected in terms of both kinematics and thermodynamics.

A poorly understood major issue is what powers the bright emission lines in central cluster galaxies. Since the ratios of X-ray/H$\alpha$, H$\alpha$/CO \cite{olivares19,tremblay18}, and H$\alpha$+[NII]/H$2$ \cite{lim12} are almost constant along the filaments, the mechanism does not seem to be one, like ionization by the AGN, that depends on the distance to the center. Most of the emission line ratios can be well described by the reprocessing of the EUV and soft X-ray radiation from the cooling plasma, as reported by several studies \cite{donahue11,donahuevoit91,polles21,ferland09,ferland08,fabian11}. In these models, which do not consider mixing layers or shocks, EUV and soft X-ray photons can create and excite the different ions that emit in the cooling gas \cite{polles21}. A large fraction of the hot cooling gas radiation is absorbed and reprocessed by the atomic and molecular gas slab. The self-irradiated X-ray excitation models, plus a small level of turbulent heating \cite{polles21}, can account for the constant ratios between the soft X-ray and H$\alpha$ found in the filaments, $\sim$3.8\,--\,6.7. \footnote{{In particular,} the models that best represent the ratios have an X-ray radiation field intensity of log10($G_{\rm Xray}$) of 1.0 and 1.6, respectively. For references the $G_{\rm X}=1$ corresponds to an input field of 3$\times$10$^{-15}$~erg s$^{-1}$~cm$^{-2}$~arcsec$^{-1}$.}

On the other hand, the emission produced by the mixing layers developed in cold clouds embedded in the X-ray halo may also lead to the formation of the observed emission lines in the filaments of cooling flow clusters \cite{begelman90,crawford92}. Turbulent mixing can generate gas
phases with intermediate temperatures and densities at constant pressure, as hydrodynamical instabilities and turbulence increase the contact surface between the hot and cold phases, enhancing mixing \cite{sun21,Fielding_2020}. In the case of stripped tails, \cite{sun21} argue that H$\alpha$-emitting tails originate from the turbulent mixing layers between the cold phase and the surrounding hot ICM. 

In addition to the H$\alpha$ and X-ray correlation discussed in this paper, it is currently unknown whether there is any correlation with the cold ($<$100~K) gas phase, which carries the bulk of the mass of the filaments. We leave the correlation with the cold molecular gas properties using ALMA (Atacama Large Millimeter Array) observations for a future study. Preliminary studies by \cite{salome06,olivares19,tremblay18,ciocan21} reveal a tight spatial and kinematical correlation between the CO emission line and the brightest H$\alpha$ emission line. However, the current ALMA observations are not sensitive enough to detect cold gas in the faintest H$\alpha$ filament. Therefore, CO detections of the diffuse gas are needed to explore any correlation with the cold gas.

\section*{Methods}\label{sec:method}

The cooling flow clusters studied in this work were selected based on two criteria: 1) the availability of optical observations from either the MUSE or SITELLE telescope, and 2) deep \textit{Chandra} observations with a large number of counts ($>$10$^{6}$ counts) within the central 100 kpc. With the exception of the Perseus and M\,87 clusters, the sources analyzed in this paper were drawn from \cite{olivares19}, however, not all clusters were included due to the lack of deep \textit{Chandra} X-ray observations needed to perform X-ray imaging decomposition analysis.

As shown in Table~\ref{table:sample}, the cooling flow clusters studied in this paper span a wide range of redshifts, $0.00428<z<0.1028$, and properties. For instance, the sample exhibits nearly a factor of two in mass deposition rate, a few in average X-ray halo temperature and star formation rate, and over 2 orders of magnitude in H$\alpha$ luminosity and 5 in total cold molecular gas mass. These strong cooling flows have been studied extensively in the literature. In Table~\ref{table:sample} we summarize some of the properties and features of each system.

We assume that $H_{\rm 0}$= 70~km~s$^{-1}$ Mpc$^{-1}$, $\Omega_{M}$ = 0.3.
The distance to the clusters is derived from the assumed cosmological parameters. At the distances of the Perseus cluster, M87, Centaurus, Hydra-A, Abell 1795, Abell 2597, PKS0745-19, 1$\arcsec$ corresponds to 0.356~kpc, 0.088~kpc, 0.208~kpc, 1.067~kpc, 1.218~kpc, 1.545~kpc, and 1.889~kpc, respectively.

\subsection*{\textit{Chandra} data analysis and observations} 
For each cluster, we used almost all the available \textit{Chandra} observations (see Supplementary Table~\ref{table:1}). The data reduction and calibration of the \textit{Chandra} observations were carried out using Chandra Interactive Analysis Observations software (CIAO) 4.13 \cite{fruscione06}, and Chandra Calibration Database (CALDB) 4.9.2.1. We performed a standard calibration, and the data was reprocessed using the chandra\_repro tool of CIAO. Standard blank sky background files and readout artifacts were subtracted.  

\subsubsection*{pGMCA decomposition method}

The decomposition method used in this study is based on the General Morphological Components Analysis (GMCA), a blind source separation algorithm introduced for X-ray observations by \cite{Bobin15}. GMCA can disentangle spectrally and spatially mixed components from an X-ray data cube, containing events of the form $(x, y, E)$, by using the sparsity of the wavelet representation of each $(x,y)$ slice of the cube. The algorithm only requires the number of components to disentangle and focuses solely on the spatial and spectral morphological diversities of said components. \cite{Picquenot19} demonstrated that the algorithm was able to extract clear, unpolluted, and physically meaningful components from highly entangled X-ray data sets. An updated version of the algorithm, pGMCA, has been developed to take into account the Poissonian nature of X-ray data \cite{Bobin20}. This version was used on Cassiopeia A data to probe the three-dimensional morphological asymmetries in the ejecta distribution  \cite{Picquenot21}. The study showed that pGMCA was perfectly suited for producing clear, detailed, and unpolluted images of both thermal and non-thermal components at different energies. 

pGMCA is an efficient method for separating the various components that contribute to the X-ray emission of cooling flow clusters, namely sloshing spirals, X-ray filaments, and cavities. However, the algorithm's performance is highly dependent on the specific case, with a large number of counts required for highly entangled datasets with overlapping components. As depicted in Supplementary Figure~\ref{fig:pGMCA_example}, when pGMCA is applied to \textit{Chandra} observations of strong cooling flow clusters, it decomposes the data into multiple components within the 3D $(x,y,E)$ space. This allows X-ray filaments at the centers of cool core clusters to be properly separated from the underlying bright hot halo and complex structures such as sloshing fronts and X-ray cavities. For Perseus cluster, we used 4 components. For M87, and Centaurus clusters we used 3 components. While for Abell\,1795, Abell\,2597, Hydra-A, and PKS0745$-$19 we used 2 components.

To apply this algorithm, we first created energy band count maps between 0.2~keV and 7~keV, using different spatial and spectral binning, depending on the depth of the \textit{Chandra} observations. We then arranged all count maps in a cube form, with energy in keV as the spectral axis. We tested different combinations of spatial and spectral (delta energy, $\Delta E$) binning. This choice depends primarily on the depth of the \textit{Chandra} observations and the physical spatial resolution of each source. The spatial and spectral binning for each source is listed in Supplementary Table~2. To better decompose the different components, we masked the central region where the AGN is located ($\sim$2--4\,arcsec). In the case of M87, the jet was also masked. 

Count rate maps of the X-ray filaments were produced for each energy band by dividing the count map cube (X-ray filament cube), obtained using the pGMCA method, by the exposure time within each energy band. The flux maps of the X-ray filaments were computed using Xspec version 12.13c and the energy conversion factor in the 0.5\,--\,2.0 keV band, which was obtained using the average X-ray temperature of the clusters within the filament radius. We used the AtomDB (version 3.0.9) database of atomic data and the solar abundance table ASPL. The ACIS-I and ACIS-S responses have varied significantly over the lifetime of \textit{Chandra}. Therefore, we computed the energy conversion factors (ECFs) for every pointing's epoch and then weight-averaged them according to the exposure time.

As shown in Supplementary Figures~\ref{fig:Ha_Xray_images} and \ref{fig:Ha_Xray_images2}, the X-ray filaments with extents of 10\,--\,70~kiloparsecs appear to be morphologically and spatially associated with the H$\alpha$ emitting filaments. The morphological relation is more robust for closer clusters, as the resolution is degraded after spatial binning of the \textit{Chandra} observations for the further clusters ($z\geq0.10$), namely, PKS0745-19. In contrast, the spatial distribution of the X-ray diffuse halo component does not correlate with the H$\alpha$ filaments. %

Previous studies have isolated X-ray filamentary structures in only Perseus \cite{sanders07} and Centaurus \cite{sanders16} clusters by fitting multitemperature components to \textit{Chandra} observations. The multitemperature method utilizes a fixed number of temperatures, but allows for varying normalization and tying together the metallicities of each component, while the pGMCA method does not required any input on the physical nature of each component. Despite that, the X-ray components (e.g., X-ray halo, sloshing spiral, and cavities), X-ray filaments and temperatures obtained using the multitemperature method are consistent with our results obtained with pGMCA method in both cases (Perseus and Centaurus, see Fig. 19 of \cite{sanders07} for an example). As previously mentioned, pGMCA also requires a large number of counts to perform the decomposition of the X-ray emission. However, for the same observations and number of counts (e.g., Perseus cluster), the multitemperature method has the drawback of degrading spatial resolution and missing faint filaments. In contrast, the pGMCA method is able to detect the faintest X-ray filaments, with a higher resolution. 

As comparison, to estimate the X-ray surface brightness of the filaments obtained through the multitemperature method, we referred to Figure~20 of \cite{sanders16} for the Centaurus cluster. Using an apec plus a phabs model, we calculated the surface brightness using the normalization presented in Figure~20 of \cite{sanders16}. We used a temperature of 0.5 keV, a metallicity of 1.2 Z$\odot$, and a galactic column density of 8.10$\times$10$^{20}$ atoms/cm$^{2}$ \cite{sanders16}. Our rough estimate of the X-ray/H$\alpha$ surface brightness ratio for one of the filaments is approximately 3.5, which is consistent with our estimates.

\subsubsection*{X-ray spectral analysis}\label{sec:x-ray-fit}

For clusters such as Perseus, Centaurus, and M\,87, where there are sufficient photons, we extracted spectra of the X-ray filaments. Since the soft X-ray line emission could be due to cooling gas, we fitted each spectrum with a model for a thermal plasma in collisional ionization equilirbrium (apec) and a galactic absorption (phabs) model of PyXspec version 12.13c \cite{PyXspec}. This allowed us to obtain the normalization, $norm$, and electron temperature, $T_{\rm e}$. We set the temperature and normalization as free parameters while fixing the abundance to 1~Z$_{\odot}$. The normalization is that obtained for each X-ray filament by fitting the X-ray spectrum with a ionized thermal plasma model (apec).
We obtained the Response Matrix Files (RMFs) and Ancillary Response Files (ARF) using the {\sc specextract} package from the observation with the highest exposure time. We tested different RMF and ARF files to fit the X-ray spectra of the filaments, and the results were not impacted.

The adopted values of the Galactic column densities ($N_{\rm H}$) were obtained from the Colden (Galactic Neutral Hydrogen Density Calculator). For Perseus, M\,87, and Centaurus we adopted a Galactic column density of 14.55$\times$10$^{20}$ atoms/cm$^{2}$, 2.54$\times$10$^{20}$ atoms/cm$^{2}$, and 8.10$\times$10$^{20}$ atoms/cm$^{2}$, respectively.

\subsection*{Optical observations and data reduction}
We used MUSE observations of six cooling flow clusters: M\,87 \cite{olivares19}, Centaurus \cite{olivares19,hamer19}, A2597 \cite{tremblay18}, Hydra-A, A\,1795 \cite{Tamhane23}, and PKS\,0745-19, as well as SITELLE observations for the Perseus cluster, to measure the optical properties of the filaments. Supplementary Table~\ref{table:MUSE} summarizes the observational properties of the MUSE and SITELLE observations for each cluster. 

\subsubsection*{MUSE observations}
MUSE is an optical integral field spectrograph installed on the Very Large Telescope (VLT). It covers a $1\arcmin\times1\arcmin$ field of view, has a pixel size of $0.2\arcsec$, and captures wavelengths ranging from 4750$\AA$ to 9350$\AA$. All the MUSE observations were taken in the wide field mode (WFM), which has a spectral resolution of $R=3000$. Observations for Hydra-A, A\,1795, A2597, and PKS0745-19 were obtained from the ESO program led by S. Hamer, while those for the Centaurus cluster were obtained from ESO programs 094.A-0859(A) and 0103.A-0447, also with S. Hamer as the principal investigator. The MUSE observation for M\,87 was obtained from ESO program 60.A-9312 (PI: Science Verification). The total exposure time varies for each source, ranging from 2700 to 11010 seconds. 
The MUSE data were processed using MUSE pipeline version 1.6.4 and ESOREX 2.1.5. These tools performed a standard procedure of reducing individual exposures and combining them into a final datacube. In addition to the sky subtraction provided by the ESOREX pipeline, we included a sky subtraction using ZAP (Zurich Atmosphere Purge; \cite{soto16}). The final data cube was then fitted using {\sc Platefit}, following the method of \cite{olivares19}. In this approach, each emission line is modeled with one Gaussian profile.
To correct for Galactic foreground extinction, we used the recalibration of the \cite{schlafly11} dust map of the Milky Way, based on IRAS+COBE data \cite{schlegel98}. We assumed a value of R$_{\rm V}$ = 3.1. Spaxels with S/N$<$7 and velocity dispersion smaller than 50~km~s$^{-1}$ were masked.
We produced spatially resolved flux maps of emission lines relevant to our study, including H$\alpha$, H$\beta$, [SII], and [NII].

\subsubsection*{SITELLE observations}

SITELLE observations were used for the Perseus cluster. SITELLE is an optical imaging Fourier transform spectrometer on the CFHT (Canada France Hawaii Telescope) that provides a field of view of 11$\arcmin\times$11$\arcmin$, a seeing-limited spatial resolution of approximately 1.0$\arcsec$, and a pixel sampling of $0.32\arcsec\times0.32\arcsec$. The observations were performed on on January 16, 2016 during the Science Verification phase (P.I. G. Morrison; \cite{gendron-marsolais18}) with the filter SN3 ($\lambda$ = 651\,--\,685 nm; $R=\lambda/\Delta\lambda\approx$ 7690).

The SITELLE software ORCS (Outils de Réduction de Cubes Spectraux, version 3, \cite{martin15}) was used for data reduction and calibration. We used the same H$\alpha$ map presented and described in \cite{gendron-marsolais18}. As described in \cite{gendron-marsolais18}, the authors fitted the data cube centered on H$\alpha$ (filter SN3) binned by a factor of 2. The spectra were fitted using a Gaussian function convolved with the instrumental line shape \cite{martin16}. The optical emission lines were fitted simultaneously with the velocity fixed by H$\alpha$ position at $\sim$5200 km/s, and the broadening of the lines was kept the same to reduce the number of free parameters. 

\subsubsection*{Data availability}
The X-ray and optical data that support the plots within this paper and other findings of this study are either publicly released (Chandra, XMM-Newton and Very Large Telescope/MUSE data) or published (narrow-band imaging data), as shown in Supplementary Tables~\ref{table:1} and~\ref{table:MUSE}. The key results of this work are also attached as an online table. Other results and reduced images of this work are available from the corresponding author V.O. upon reasonable request.  \href{https://doi.org/10.6084/m9.figshare.27760053}{Source data} are provided with this paper . 

\subsubsection*{Code availability}
The software to reduce and analyze the X-ray and optical data in this work is publicly released. Upon request, the corresponding author V.O. will provide the code (Python) used to produce the figures.

\subsubsection*{Acknowledgements}
V.O. acknowledges support from DICYT through grant 1757 Comité Mixto-ESO Chile, and NASA NPP funding. V.O. and Y.S. were supported by NSF grant 2107711, Chandra X-ray Observatory grant GO1-22126X, GO2-23120X, and NASA grant 80NSSC21K0714. 
M.G. acknowledges support from the ERC Consolidator Grant \textit{BlackHoleWeather} (101086804). 
M.L.-G.M. acknowledges financial support from the grant CEX2021-001131-S funded by MCIU/AEI/ 10.13039/501100011033, from the coordination of the participation in SKA-SPAIN, funded by the Ministry of Science, Innovation and Universities (MCIU), as well as NSERC via the Discovery grant program and the Canada Research Chair program.

This research has made use of software provided by the \textit{Chandra} X-ray Center (CXC) in the application packages CIAO. 
Based on observations collected at the European Organisation for Astronomical Research in the Southern Hemisphere under ESO programme(s): 60.A-9312(A), 0103.A-0447(A), 094.A-0859(A).

\subsubsection*{Author Contributions}
V.O. initiated the research, reduced and analyzed the \textit{Chandra} and MUSE observations, and wrote the manuscript. A.P. assisted with the \textit{Chandra} analysis and performed the pGMCA decomposition. Y.S. aided in analyzing the \textit{Chandra} observations. M.G. conducted the analysis of the simulations. A.P., Y.S., and G.M. contributed to writing the manuscript. M.G-M. provided the SITELLE H$\alpha$ data of the Perseus cluster. All authors contributed to the discussion and interpretation of the results.

\subsubsection*{Correspondence} Correspondence and requests for materials should be addressed to Valeria Olivares (valeria.olivares@usach.cl).

\subsubsection*{Ethics declarations Competing interests}
The authors declare no competing interests.

\subsubsection*{Competing interests} The authors declare that they have no competing financial interests.

\section*{Tables}

\begin{table}[h]
\caption{Properties of the sample}\label{table:sample}
\begin{tabular}{ccccccccc} 
\toprule
Cluster  & Redshift & kT  & SFRs & $\dot{M}_{\rm cool}$& $M_{\rm mol}$ & H$\alpha$ luminosity  \\ 
 & & (keV) & ($\rm M_{\odot}$yr$^{-1}$) &  ($\rm M_{\odot}$~yr$^{-1}$) & ($\rm M_{\odot}$) &  (erg~s$^{-1}$) &   \\ 
 (1) & (2) &(3) & (4) &  (5) &  (6) & (7)\\ 
\midrule
Perseus  & 0.01756 &6.79& 30$\pm$23 & 467.7$\pm$1.12 & 8.5$\times$10$^{10}$ & 5.0$\times$10$^{41}$\\
M87 & 0.00428 & 2.50 &   1.3 &  19.5$\pm$1.00  & 4.7$\times$10$^{5}$  & 9.7$\times$10$^{39}$ \\
Centaurus  & 0.01041 &  3.96 &  0.2$\pm$0.1 &  9.3$\pm$1.00   & 0.9$\times$10$^8$ & 2.7$\times$10$^{40}$ \\
Hydra$-$A & 0.05490 &4.30& 8$\pm$7 &    109.6$\pm$1.04  & 5.4$\times$10$^{9}$ &2.1$\times$10$^{41}$\\
Abell\,2597  & 0.08210 &3.58 & 5$\pm$8 & 309.0$\pm$1.12 & 2.3$\times$10$^{9}$ & 1.1$\times$10$^{42}$\\
Abell\,1795 & 0.06330 & 7.80 &  8$\pm$8 & 186.2$\pm$1.04& 3.2$\times$10$^{9}$ &3.9$\times$10$^{42}$\\
PKS0745$-$19 & 0.10280 & 8.50 & 70$\pm$94 &776.2$\pm$1.04 & 4.9$\times$10$^{9}$ & 3.2$\times$10$^{42}$\\
\hline
\end{tabular}
\footnotetext[1]{Cluster name.}
\footnotetext[2]{Redshift.}
\footnotetext[3]{Average cluster temperature (ACCEPT catalog) \cite{cavagnolo09}.}
\footnotetext[4]{ Star formation rate - \cite{mittal15}, and for M\,87 \cite{McDonald_2018}. }
\footnotetext[5]{Mass deposition rate taken from \cite{mcdonald18}, calculated inside the radius where the cooling time is less than 3 Gyr.}
\footnotetext[6]{Cold molecular gas mass - \cite{salome03,edge02,salome06,salome08,olivares19,russell19}.}
\footnotetext[7]{Total H$\alpha$ luminosity (this work).}
\end{table}

\bibliography{main_paper_v2_main.bbl}

\clearpage
\begin{appendices}

\section*{Supplementary Information}\label{sec:supplementary}

\subsection*{X-ray and H$\alpha$ surface brightness and luminosity of each filament}\label{sec:regions}
Filament regions were selected visually using the H$\alpha$ maps for each cluster, following coherent structures spatially and kinematically. The {\sc CIAO} package {\sc dmextract} was used to obtain the surface brightness for each filamentary region in the X-ray and H$\alpha$ emitting nebulae. The H$\alpha$ and X-ray luminosities, $L$, of each region were calculated as follows:
\begin{equation}
L = f \times 4\pi D_{\rm L}(z)^{2}, 
\end{equation}
in units of erg~s$^{-1}$, where $f$, is the X-ray or H$\alpha$ flux of each region, and $D_{L}(z)$ is the redshift-dependent luminosity distance. 

Following \cite{sun21}, the surface brightness for both X-ray and H$\alpha$ components is defined as the luminosity surface brightness, in units of erg\,s$^{-1}$\,kpc$^{-2}$. This is calculated by dividing the total luminosity by the physical area of the region in kpc$^{2}$. 

We used the {\sc linmix} Bayesian method \cite{Kelly07} to fit the X-ray -- H$\alpha$ surface brightness correlation for all the filament regions in our sample. A major advantage of using Bayesian formalism for linear regression is that the scatter is treated as a free parameter, along with the normalization, intercept and slope \cite{gaspari19}. The best-fit of the correlation is $SB_{\rm X-ray} = (3.44\pm1.08)~SB_{\rm H\alpha}^{(0.94\pm0.08)}$ ($\rm log10(SB_{\rm X-ray})= (0.53\pm0.03)+(0.94\pm0.08)log10(SB_{\rm H\alpha})$). The units of both $SB_{\rm X-ray}$ and $SB_{\rm H\alpha}$ are $\rm 10^{38} \, erg\, s^{-1}\, kpc^{-2}$. Each parameter corresponds to the average of the distributions with 1$\sigma$ errors given by the standard deviation. If we exclude the upper limit from the fit and only include detected X-ray filaments, the best fit is almost the same as the one with upper limits: $SB_{\rm X-ray} = (3.44\pm1.07)~SB_{\rm H\alpha}^{(0.90\pm0.07)}$. If upper limits are included, then the intrinsic scatter is 0.06$\pm$0.01 in log10 scale (between 0.04 and 0.10), and the correlation coefficient is 0.80$\pm$0.04. If we include only detections (excluding upper limits), the intrinsic scatter of the relation is 0.06$\pm$0.01 in log10 scale (between 0.03 and 0.12), and the linear correlation coefficient is 0.80$\pm$0.04. 

We investigated whether the X-ray/H$\alpha$ ratios change depending on the distance from the central galaxy, as expected from ionizing mechanisms that heavily rely on the central AGN, such as photoionization from the central AGN and shocks. According to \cite{sun21}, continuous mixing between the surrounding hot halo and the warm gas should deplete the cold gas, leading to a lower X-ray/H$\alpha$ at large radii. The supplementary figure~\ref{fig:Xray_Ha_distance_size} shows that there is no significant dependence of the X-ray/H$\alpha$ ratio on distance, over a length of $\sim$40 kiloparsecs. Additionally, we found no correlation between the X-ray/H$\alpha$ SB ratios and the size of the region in kpc$^{2}$, which indicates our results are not affected by our choice of region size. 

The scatter in the X-ray--H$\alpha$ surface brightness correlation is multifold. From 2D X-ray/H$\alpha$ ratio maps (see Supplementary Figure~\ref{fig:2D_ratio_maps}), we observe that the ratio is significantly higher (X-ray/H$\alpha \geq 6$) in areas where the H$\alpha$ filaments are more diffuse or located at the edges of the X-ray filaments. This is likely due to residual emission from the X-ray halo. In contrast, the X-ray/H$\alpha$ ratio is lower (X-ray/H$\alpha \leq4$) in areas where the H$\alpha$ gas is clumpier and more compact.

Additionally, the X-ray and H$\alpha$ data have different depths and resolutions. The X-ray observations of more distant clusters in the sample, such as Hydra-A, Abell\,1795, and PKS 0745-19, are much shallower than those of other clusters such as Centaurus and Perseus. In such cases the GMCA algorithm is less effective at separating the X-ray filaments from the bright X-ray halo, leaving behind residual emission and unresolved X-ray filaments due to the high spatial binning required to provide reliable X-ray imaging decomposition.

{Finally, in Centaurus, one of the filaments is not detected in the X-ray filament image and, thus, it only provides an upper limit.}

\subsubsection*{Comparison with CCA simulations} \label{sec:CCA}
In figure~1, we showed the predicted correlation between X-ray and H$\alpha$ surface brightness predicted from CCA by leveraging the results of hydrodynamical simulations (see \cite{gaspari17} for the numerical details). 
We computed the surface brightness by integrating the emissivity along line of sight ($\approx n^2 \Lambda$) for the thermodynamic profiles of the CCA runs. In particular, during the CCA rain, the radial density profiles of the different phases are highly correlated with a logarithmic slope of -1.

In keeping with the observational analysis, we used the soft X-ray band ($T\sim10^7$\,K) to compute the cooling function, which is fairly constant at $\Lambda \sim 3\times10^{-23}$ erg\,s$^{-1}$~cm$^{3}$. For the H$\alpha$ emission we use the hydrogen recombination emissivity $\Lambda \sim 4\times10^{-25}$ erg\,s$^{-1}$~cm$^{3}$ \cite{gaspari15}, with the simulated warm gas stabilizing at a temperature of $10^4$ K (the first floor of the top-down condensation cascade, before reaching the molecular phase regime). 
We iterated the projection along the line of sight by randomly varying the impact parameter 100 times. Thus, the simulated correlation accounts for uncertainties due to projection, alongside the scatter of the thermodynamic profiles and filling factor. The latter is typically of the order of $10^{-1}$ and $10^{-3}$, for the X-ray and H$\alpha$ filaments respectively.

As for the observational analysis, we used the Bayesian {\sc linmix} method \cite{Kelly07} to robustly fit the CCA X-ray and H$\alpha$ surface brightness points, which also provides robust uncertainties on all the fitting parameters. For the CCA points we retrieve the following correlation: $\log {\rm SB_x} = (0.54\pm0.02) + (0.98 \pm 0.04) \log {\rm SB_{H\alpha}}$. As shown in Figure~1 these parameters agree well with the observed values, both in terms of normalization and slope.
The simulated intrinsic scatter is $0.19\pm0.09$ in log scale ($55\%$), moderately larger than the observed data, although still consistent within the 1-RMS uncertainties. This is likely due to the ability of simulations to trace the multiphase gas over a wider range of brightness. In the future, it will be key to increase the detections of observed filaments also at the very low- and high-end regimes.

\subsection*{Properties of the X-ray and optical filaments}\label{sec:properties_fil}

\subsubsection*{Properties of the X-ray filaments}
For the X-ray filaments, we estimated the electron density ($n_{\rm e}$) using the normalization ($norm$) values obtained for filamentary regions with high counts in the three cooling flow clusters with higher spectral resolution: Perseus, Centaurus, and M\,87 clusters. The estimation is done as follows:
\begin{equation}
n_{\rm e}^{2} = \frac{4\pi~10^{14}~(1+z)^{2}~D_{\rm A}^{2}~norm}{\rm (0.85~V)}, \label{eq:2}
\end{equation}

The estimation uses equation~\ref{eq:2}, where $n_{\rm e}$ is in $\rm cm^{-3}$, $D_{\rm A}$ is the angular distance to the source in cm, and $V$ is the volume of the emitting region in $\rm cm^{3}$. We assumed a cylindrical geometry for the filaments, with a volume defined as $\rm V = \pi\,l (w/2)^{2}$, where $\rm w$ and $\rm l$ are the projected width and length of each filament region. In our work, we measure the width and length of each filament from the optical observations, since the X-ray data have lower resolution. Still, the optical MUSE and SITELLE observations are not spatially resolved at the filament width scales. High spatial resolution Hubble (HST) observations resolved optical filaments in Perseus and Centaurus to have widths of about 75~pc and 50~pc, respectively \cite{fabian08, fabian16}, while the seeing of the MUSE and SITELLE observations is 70~pc for M\,87, 200~pc for Centaurus, and 350~pc for Perseus. We expect the X-ray filaments to be relatively wider than the H$\alpha$ filaments detected by HST as they are hotter, and therefore more diffuse. Using a cylindrical geometry for the X-ray filaments, we obtained values for the electron density of the X-ray filaments between 0.02 to 0.45~cm$^{-3}$, with an average value of 0.15~cm$^{-3}$.

Another way to estimate the volumes of X-ray filaments is to assume that they are in thermal pressure equilibrium with the ambient hot X-ray halo \cite{sanders07, werner13}. We compared the electron gas pressure of the X-ray filaments to the deprojected X-ray profile of the X-ray halos from the ACCEPT catalog \cite{cavagnolo08}. For M\,87, the X-ray halo pressure ranges from 4$\times$10$^{-10}$\,dyn\,cm$^{-2}$ to 1$\times$10$^{-10}$\,dyn\,cm$^{-2}$ at distances of 1 and 6 kpc from the cluster center, respectively (see also \cite{churazov08}). The estimated emitting volume of the X-ray filaments is 3$\times$10$^{62}$ to 3$\times$10$^{63}$ cm$^{3}$ (0.01\,--\,0.07 kpc$^{3}$), corresponding to widths of 0.1\,--\,0.3 kpc. These volumes are 1 to 4 times smaller than those measured for the optical filaments.

In the case of the Centaurus cluster, the deprojected X-ray halo pressure ranges from 2.4$\times$10$^{-10}$\,dyn\,cm$^{-2}$ to 1.4$\times$10$^{-10}$\,dyn\,cm$^{-2}$ over distances of 1 to 4 kpc (see also \cite{sanders16}). This results in X-ray filament volumes of 2$\times$10$^{63}$ cm$^{3}$ to 7$\times$10$^{64}$ cm$^{3}$ (0.1\,--\,2.5 kpc$^{3}$), corresponding to widths of 0.3 to 0.8 kpc. Similar to M87, these volume values are about 1 to 2 times smaller than the estimated optical filament volumes, estimated based on their visible extents.

The electron pressure of the X-ray halo in the Perseus cluster ranges from 2$\times$10$^{-10}$\,dyn\,cm$^{-2}$ to 1.7$\times$10$^{-10}$\,dyn\,cm$^{-2}$ over distances of 1 kpc to 35 kpc, as reported in \cite{sanders04}. Assuming thermal pressure equilibrium between the halo and X-ray filaments, the volumes of the filaments range from 7.3$\times$10$^{63}$ cm$^{3}$ to 10$^{65}$\,cm$^{3}$ (0.2\,--\,4.0 kpc$^{3}$), corresponding to widths of 0.1\,--\,1 kpc. These volumes are 1 to 30 times smaller than those measured from optical images for the Perseus cluster. In \cite{sanders07}, the authors estimated the volume of a northern filament in Perseus using the emission measure of the filament component and the electron pressure of the surrounding medium. They found a volume of 10$^{63}$\,cm$^{-3}$. In contrast, we obtained a volume of 8$\times$10$^{63}$\,cm$^{-3}$ for the same filament region. We found that we are overestimating the width of the filaments due to spatial resolution limits when using a constant width of $\sim$150\,pc. Our results indicate that the electron pressure of the X-ray filaments increases by a factor of a few if we use the volumes obtained assuming pressure balance between the X-ray halo and X-ray filaments.

Using the electron temperature, and density of each filament, estimated using the width and length measured from the optical filaments, we derived the electron pressure, $P_{\rm e, Xray, filaments}$, of the X-ray filaments as follows:
\begin{equation}
P_{\rm e} = k\, n_{\rm e}\, T_{\rm e},
\end{equation}
where $k$ is the Boltzmann constant. We obtained values between 3$\times$10$^{-11}$~dyne~cm$^{-2}$ and 5.4$\times$10$^{-10}$~dyne~cm$^{-2}$, with an average value of 1.5$\times$10$^{-10}$ dyne~cm$^{-2}$. As stated, these values could increase by a few fold due to the inability to resolve the filaments.

Projection effects could give a smaller volume of the filaments, as some could be tilted with respect to the plane of the sky, causing the electron density to be overestimated and, consequently, the electron pressure. Assuming the filaments are positioned with a 45$\deg$ angle with respect of the sky, then, the $n_{\rm e}$, and consequently, the $P_{e}$ decrease by a factor of 1.2.\\

\subsubsection*{Properties of the H$\alpha$ filaments}

We used PyNeb \cite{pyneb} to compute the temperature and electron density of the optical filaments in the M\,87 and Centaurus clusters. For this, we used the forbidden sulfur lines, i.e., [S\,II]$\lambda$6717 and [S\, II]$\lambda$6732, as well as the auroral and nebular Nitrogen lines of each filamentary region, i.e., [N\,II]$\lambda$5755, [N\,II]$\lambda$6548, and [N\, II]$\lambda$6583. For filaments where the [NII\,5755] auroral emission lines were undetected, we assumed an electron temperature of 10$^{4}$~K to derive the electron density.

The electron temperature of the warm filaments, $T_{\rm e, optical}$, is almost constant across the nebulae with a value of 10,000~K. Meanwhile, the electron density, $n_{\rm e, optical}$, peaks at the center with values of 500~cm$^{-3}$ at the AGN position. On the warm filaments, the electron density, $n_{\rm e, optical}$, is lower between 40\,--\,150~cm$^{-3}$, with a mean value of 60~cm$^{-3}$. It is worth noting that the [S II] line ratios are sensitive to the density higher than 30~cm$^{-3}$. They become ineffective and can only provide upper limits for lower densities.

We derive the electron pressure, $P_{\rm e, optical, filaments}$, for the warm phase and obtain values ranging from 5$\times$10$^{-11}$~dyne~cm$^{-2}$ to 3.5$\times$10$^{-10}$~dyne~cm$^{-2}$ for the optical filaments in Centaurus and M87. The average value of electron pressure is 10$^{-10}$~dyne~cm$^{-2}$.

\subsubsection*{Radial dependence in the fraction of the total X-ray emission arising from the filaments?}\label{sec:halo_contribution}

As expected, cooling flow clusters with bright X-ray cores exhibit a significant radial dependence on the fraction in X-ray emitting filaments and the X-ray halo. Closer to the central region of the clusters, within 1~kpc, the fraction of X-ray emitting filaments ($SB_{\rm Xray, filaments}/SB_{\rm Xray, total}$) is about 10\%--20\%, while at greater distances (10~kpc), the fraction rapidly decreases to 1\%. For comparison, the contribution to the X-ray emission from the stripped X-ray tails with length of 20-100~kpc lies in the range 10--30\% (see Supplementary Figure~\ref{fig:SB_XrayHalo_Halpha_filaments}).

\subsection*{Filling factor of the warm gas phase}

By utilizing the physical estimates presented in this work, including electron density, temperature, pressure, as well as surface brightness of X-ray and optical emitting filaments, we can estimate the filling factor for the warm ionized gas. We do this by assuming the filling factor of the X-ray-emitting gas is equal to one.

The following equation~\ref{eq:4} is used to analyze a fixed surface area in the tail (see \cite{OsterbrockFerland2016, sun21}). The calculation assumes uniform densities for both the hot and warm gas, and that the hot gas and warm gas have the same mean molecular weight, and that the hot gas and warm gas have the same abundances in every filament.

\begin{equation}
\frac{L_{\rm Xray}}{L_{\rm H\alpha}} = 89 ~ \left(\frac{n_{\rm e,Xray}}{n_{\rm e,H\alpha}} \right)^{2} ~ \frac{f_{\rm Xray}}{f_{\rm H\alpha}}, \label{eq:4}
\end{equation}

Here, $L_{\rm Xray}$ and $L_{\rm H\alpha}$ represent the X-ray and H$\alpha$ luminosities of a given filament, respectively. Similarly, $n_{\rm e,Xray}$ and $n_{\rm e,H\alpha}$ denote the electron densities of the X-ray and H$\alpha$ filaments, while $f_{\rm Xray}$ and $f_{\rm H\alpha}$ are the volume filling factors for the hot and warm gas in the filament.

We obtained an average X-ray/H$\alpha$ surface brightness luminosity ratio of 4.1. Assuming the X-ray filaments are a volume-filling gas, with $f_{\rm Xray}$ equal to unity, and using the electron density estimates, we obtain a mean filling factor of approximately 9$\times$10$^{-4}$ for the H$\alpha$ emitting filaments.

\renewcommand{\figurename}{Supplementary Figure}
\renewcommand{\tablename}{Supplementary Table}


\begin{figure}
\centering
    \includegraphics[width=0.99\textwidth]{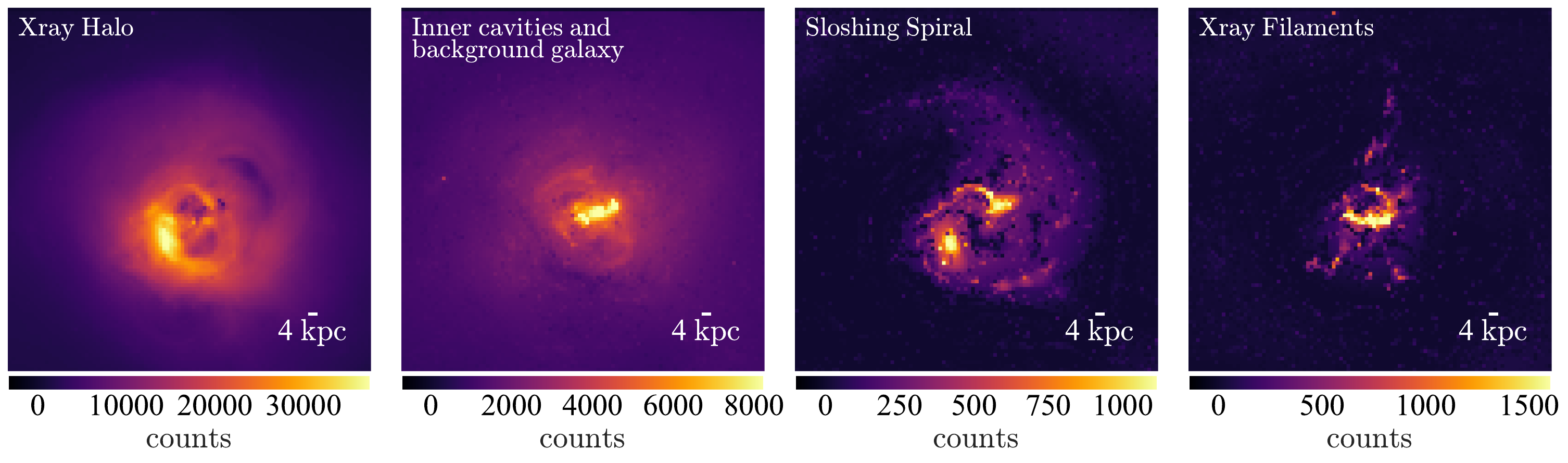}
\caption{ - Example of the pGMCA imaging decomposition method for the Perseus cluster. From left to right the components are: X-ray Halo, inner cavities plus the background galaxy, sloshing spiral, and X-ray filaments.\label{fig:pGMCA_example} }
\end{figure}

\begin{figure}
\centering
    \includegraphics[width=0.99\textwidth]{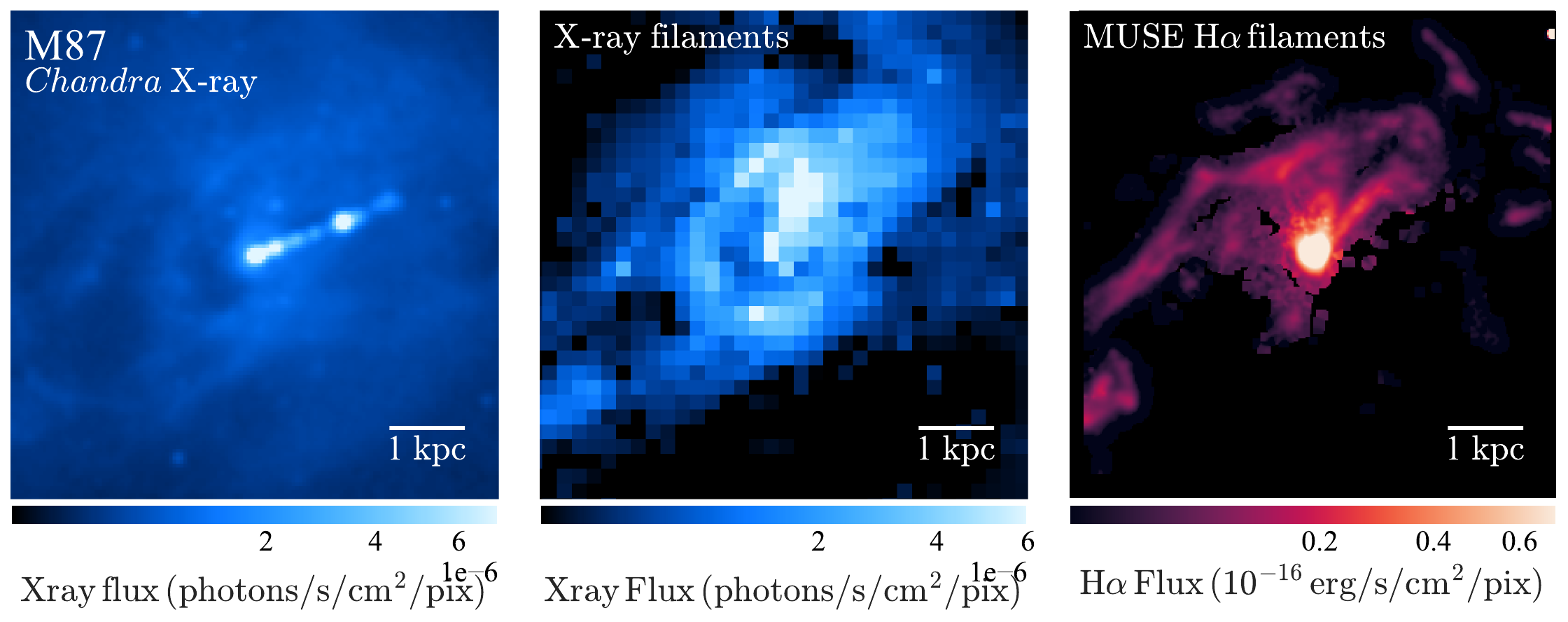}
    \includegraphics[width=0.99\textwidth]{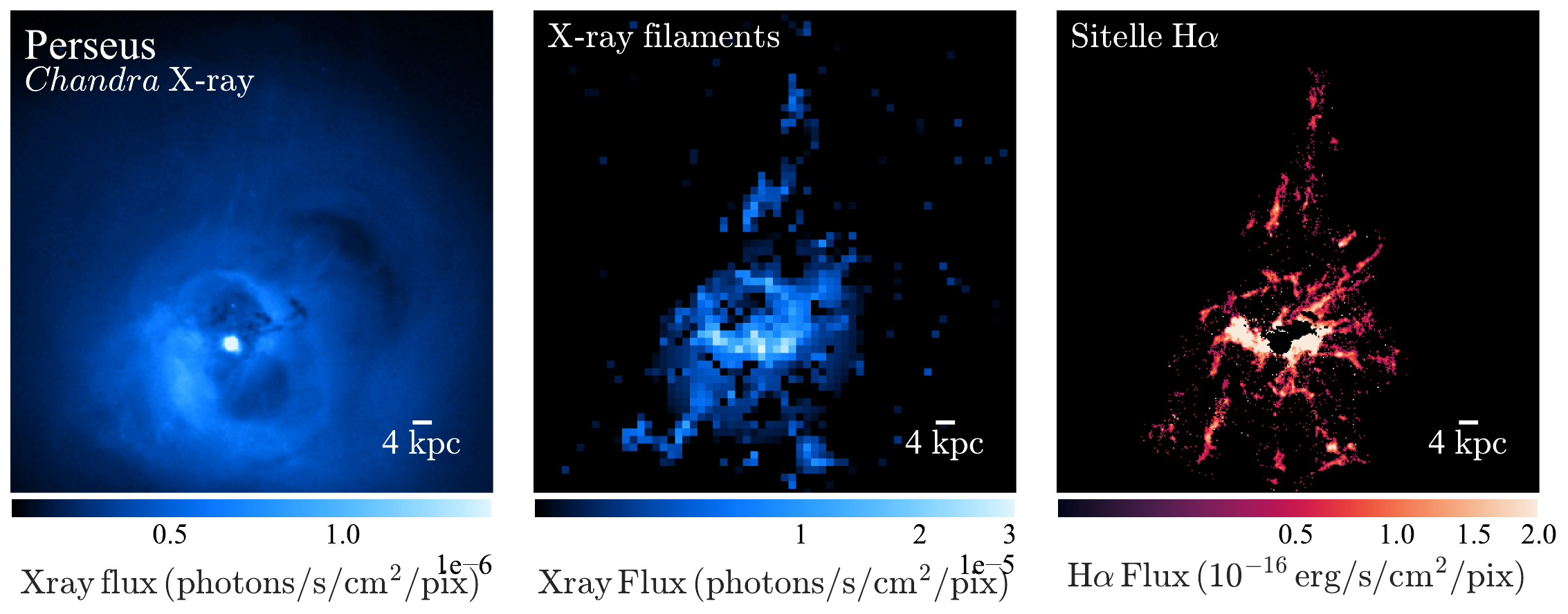}
    \includegraphics[width=0.99\textwidth]{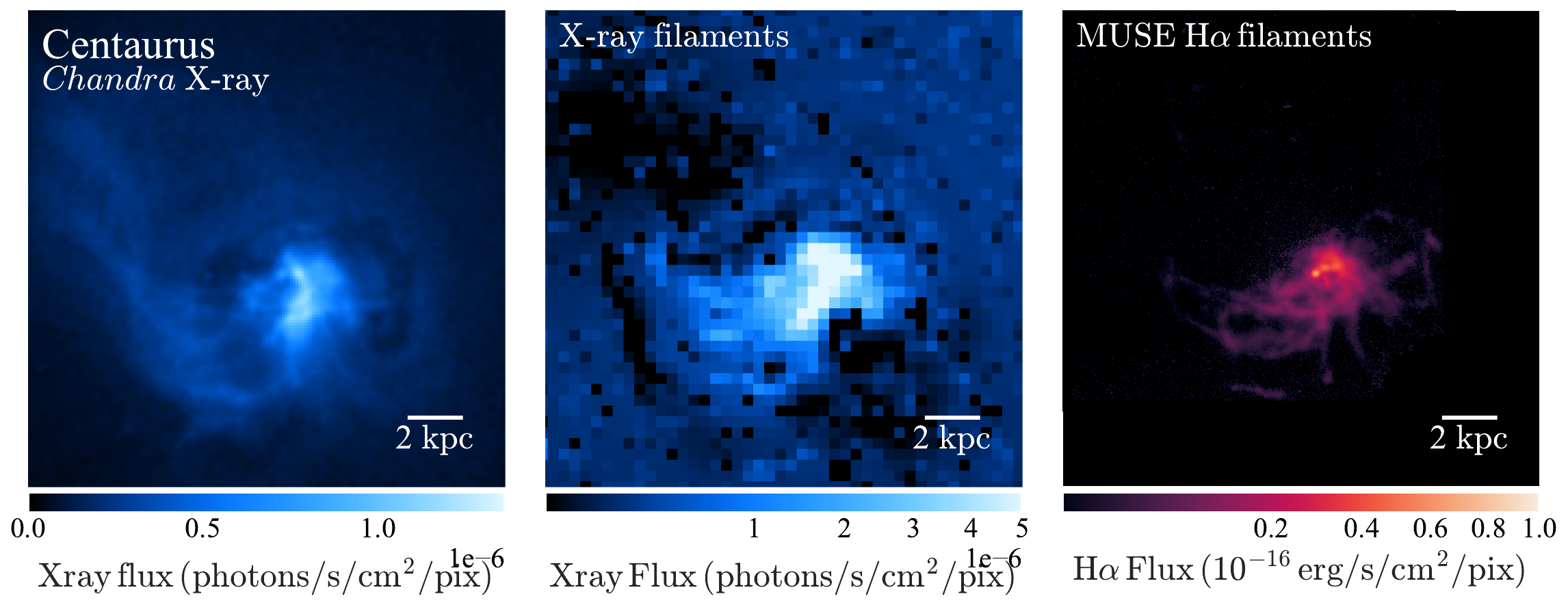}
    \includegraphics[width=0.99\textwidth]{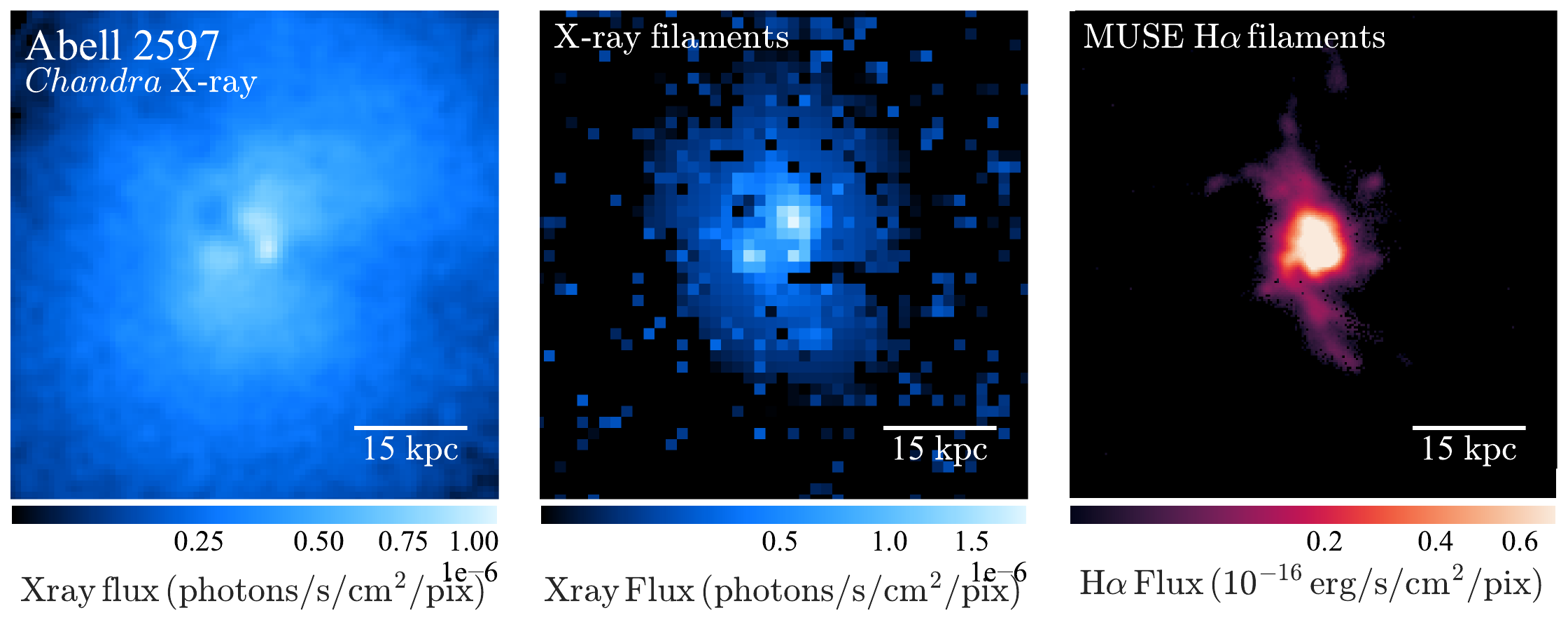}
\caption{ - Comparison of the different temperature phases of the gas for our sample. X-ray image of the cooling flow clusters from \textit{Chandra} observations (Left panel), the X-ray filaments obtained using imaging decomposition (Middle panel), and the H$\alpha$ filaments (Right panel).} \label{fig:Ha_Xray_images}
\end{figure}

\begin{figure}
\centering
    \includegraphics[width=0.99\textwidth]{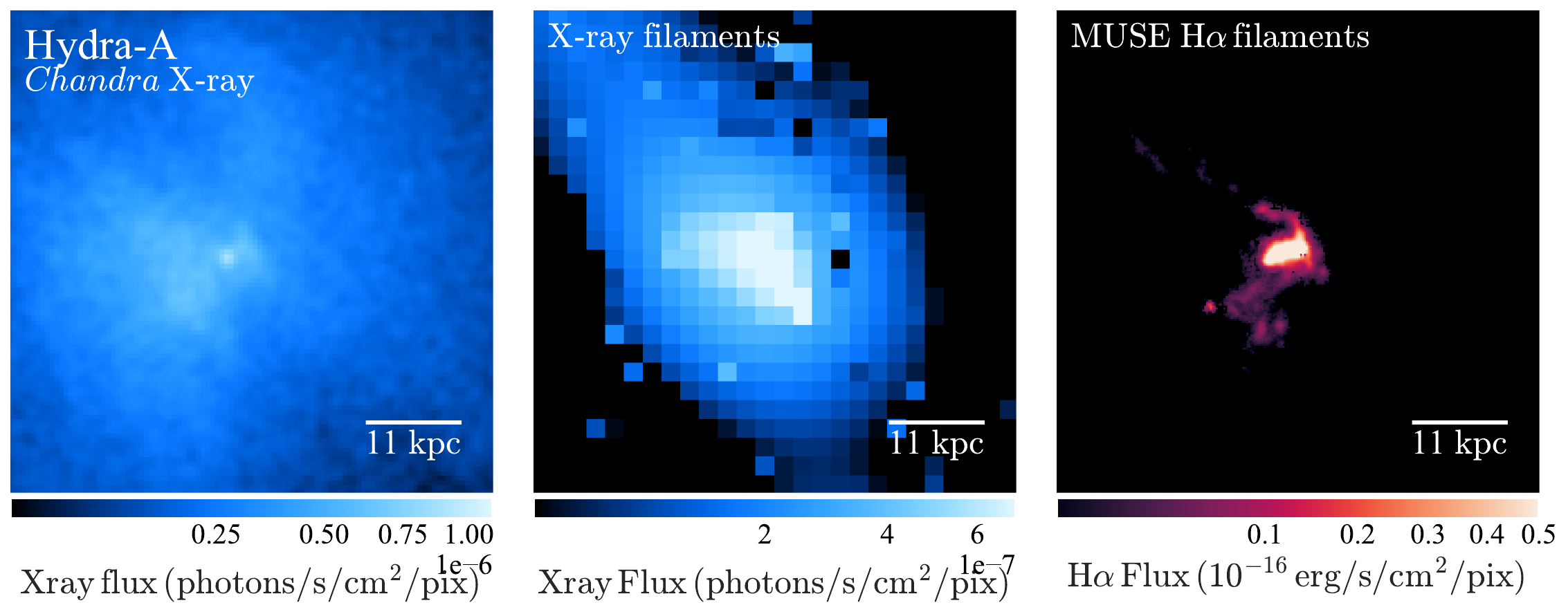}
    \includegraphics[width=0.99\textwidth]{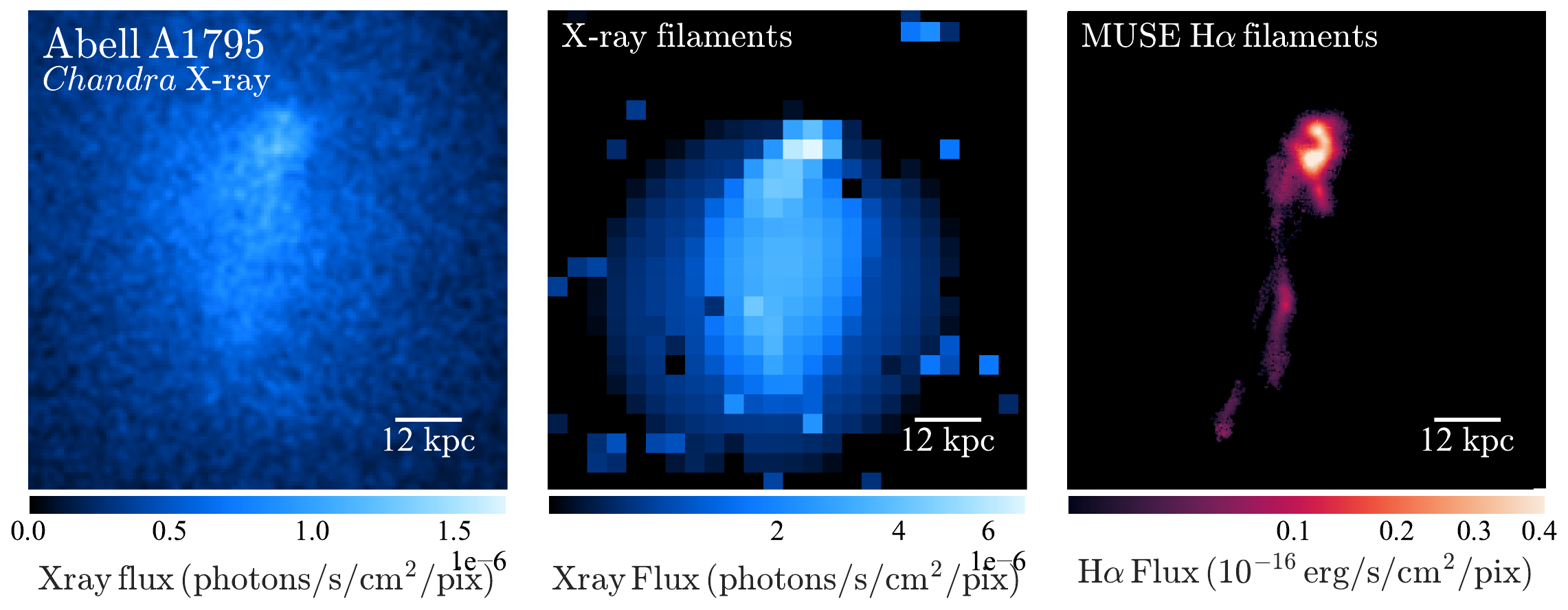}
    \includegraphics[width=0.99\textwidth]{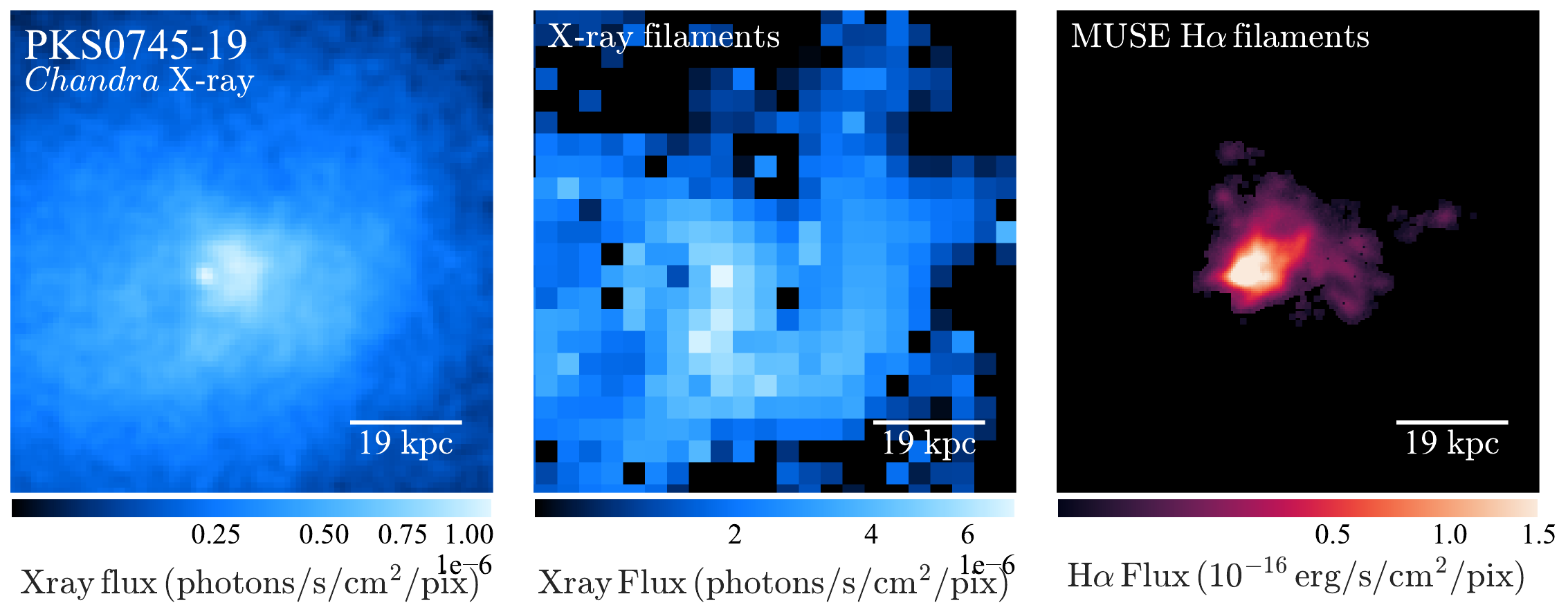}
\caption{ - Comparison of the different temperature phases of the gas for our sample. X-ray image of the cooling flow clusters from \textit{Chandra} observations (Left panel), the X-ray filaments obtained using imaging decomposition (Middle panel), and the H$\alpha$ filaments (Right panel).} \label{fig:Ha_Xray_images2}
\end{figure}

\begin{figure}
\centering
    \includegraphics[width=0.48\textwidth]{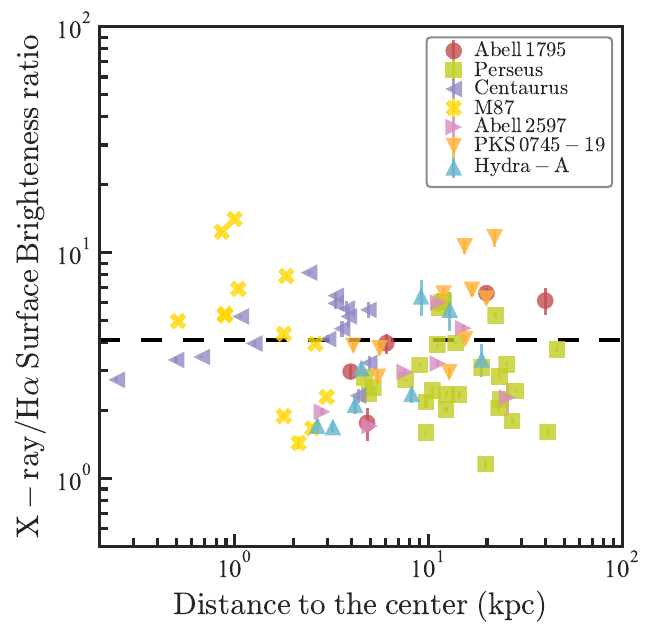}
    \includegraphics[width=0.468\textwidth]{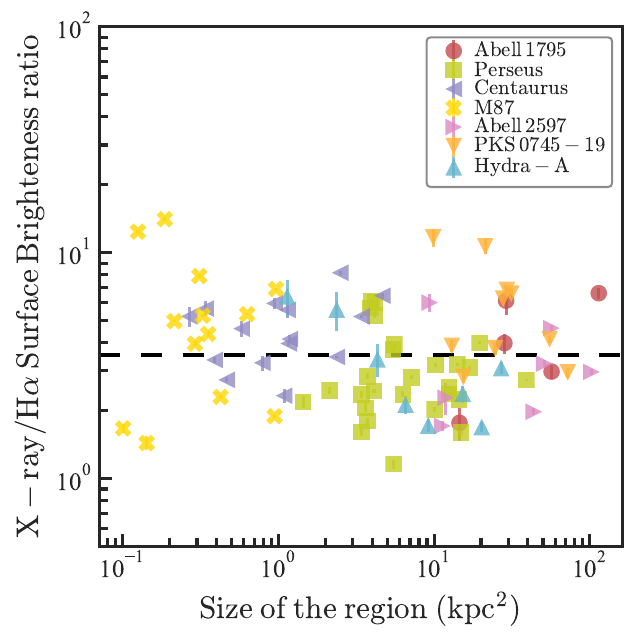}
\caption{ - X-ray/H$\alpha$ surface brightness ratio as a function of distance and size of regions. X-ray/H$\alpha$ surface brightness ratios for seven strong cooling flow clusters as a function of distance to the central galaxy (Left panel) and of the size of the filaments (Right panel). Errors are 1$\sigma$. The dashed black line corresponds to the average X-ray/H$\alpha$  surface brightness ratio of the filaments. \label{fig:Xray_Ha_distance_size} }
\end{figure}

\begin{figure}
\centering
    \includegraphics[width=0.23\textwidth]{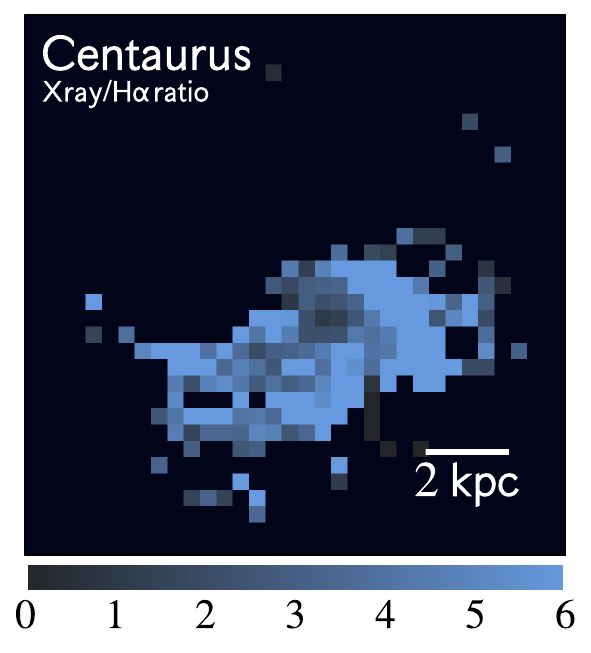}
    \includegraphics[width=0.23\textwidth]{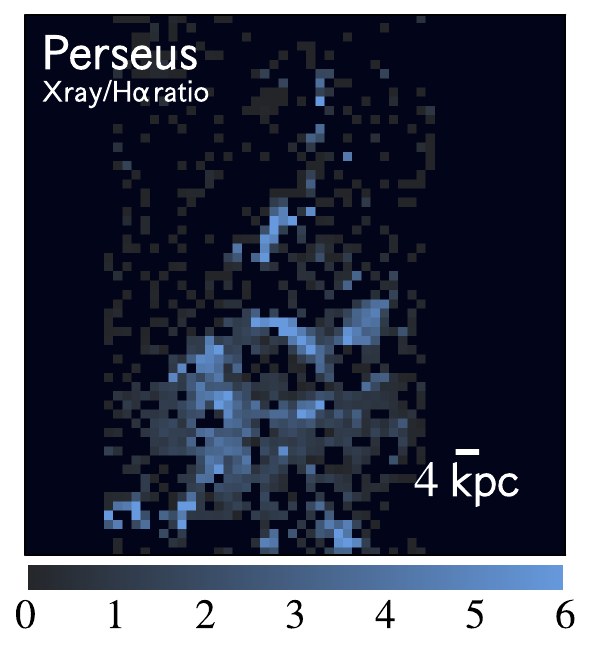}
    \includegraphics[width=0.23\textwidth]{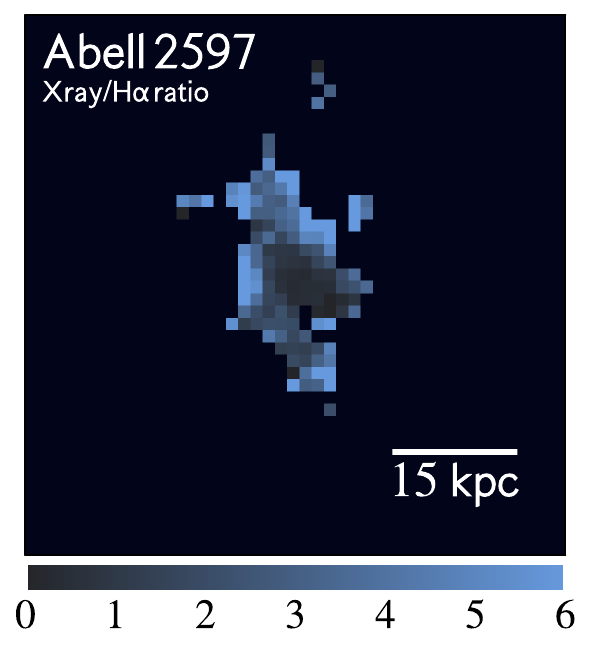}
    \includegraphics[width=0.23\textwidth]{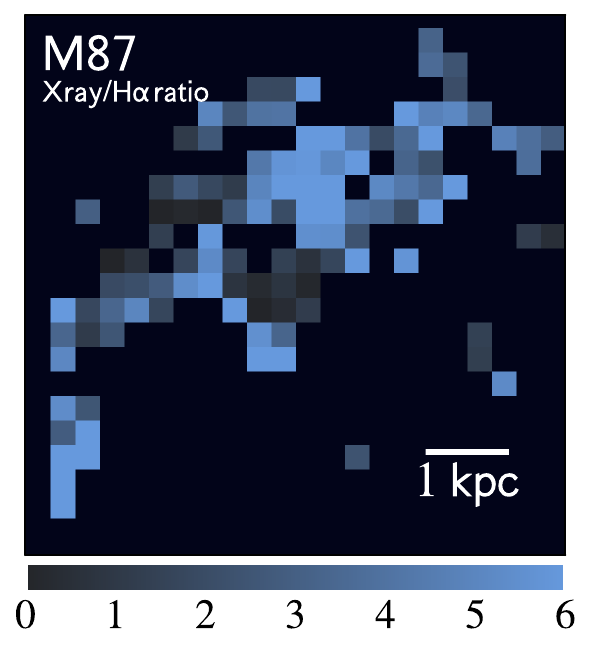}
\caption{ - Examples of X-ray/H$\alpha$ surface brightness ratio maps for Perseus, Centaurus, Abell\,2597, and M87. The H$\alpha$ maps have been projected to the spatial resolution of the GMCA map obtained from the \textit{Chandra} observations. 
\label{fig:2D_ratio_maps} }
\end{figure}

\begin{figure}
    \centering
    \includegraphics[width=0.9\textwidth]{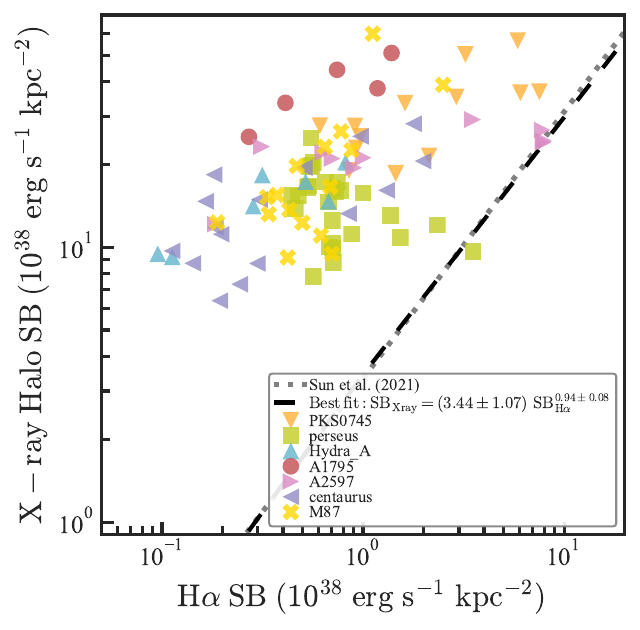}
\caption{  - Comparison of the H$\alpha$ and total X-ray surface Brightness. H$\alpha$ -- X-ray surface brightness for filaments measured from the original \textit{Chandra} X-ray images without filtering in 7 strong cooling flow clusters. Each data point corresponds to a filament region for a given cooling flow cluster. Errors are 1$\sigma$. X-ray surface brightness was computed within the 0.5\,--\,2.0 keV band using the original X-ray image. The black dashed line corresponds to the best fit of the H$\alpha$-X-ray surface brightness correlation for the filaments. The dotted gray line shows the relation found for diffuse gas in stripped tail \cite{sun21}. 
} 
\label{fig:SB_XrayHalo_Halpha_filaments}
\end{figure}

\newpage\clearpage
\begin{table}
\caption{List of {\sl Chandra} observations used in this paper.}
\centering
\begin{tabular}{c c c} 
 \hline\hline
 Cluster   & OBSID  & Total Exposure \\ 
 \hline
 Perseus & 428, 1513, 4950, 19568, 4951, 19913, 4952, 19914, 4953, 19915& 1554\,ksec \\ 
        &  502, 3209, 503, 3404, 6139, 4289, 6145,4946, 6146, 4947 & \\
        & 4949, 11713, 11714, 11715, 11716, 12025, 12033, 12036, 12037 & \\
\hline
 M87  & 241, 351, 352, 517, 1808, 2707, 3084, 3085, 3086, 3087, 3088 &  1745\,ksec\\
   & 3717, 3975, 3976, 3977, 3978, 3979, 3980, 3981, 3982, 4917&\\
   &  4918, 4919, 4920, 4921, 4922, 4923, 5737, 5738, 5739, 5740&\\
   &  5741, 5742, 5743, 5744, 5745, 5746, 5747, 5748, 5826, 5827&\\
   &  5828, 5829, 6136, 6137, 6186, 6299, 6300, 6301, 6302, 6303&\\
   &  6304, 6305, 7210, 7211, 7212, 7348, 7349, 7350, 7351, 7352&\\
   &  7353, 7354, 8047, 8057, 8063, 8510, 8511, 8512, 8513, 8514&\\
   &  8515, 8516, 8517, 8575, 8576, 8577, 8578, 8579, 8580, 8581&\\
   &  10282, 10283, 10284, 10285, 10286, 10287, 10288, 11512, 11513&\\
   &  11514, 11515, 11516, 11517, 11518, 11519, 11520, 11783, 13515&\\
   &  13964, 13965, 14973, 14974, 16042, 16043, 17056, 17057, 18232&\\
   &  18233, 18612, 18781, 18782, 18783, 18809, 18810, 18811, 18812&\\
   &  18813, 18836, 18837, 18838, 18856, 19457, 19458, 20034, 20035&\\
   &  20488, 20489, 21075, 21076, 21457, 21458, 21700, 21701, 23669, 23670 &\\
   \hline
 Centaurus &  504 ,505 ,1560 ,4190, 4191, 4954, 4955 ,5310, 16223, 16224 &778\,ksec \\ 
   &  16225 ,16534 ,16607,16608, 16609, 16610 & \\ 
\hline
 Abell 2597 &  922, 6934, 7329, 19596, 19597, 19598, 20626, 20627 & 625\,ksec \\ 
   &   20628, 20629, 20805, 20806, 20811, 20817  & 625\,ksec \\ 
\hline
 Abell 1795 &19878, 20652, 21840, 19868, 10900, 10898, 10901, 18424, 18434, 12028 & \,ksec \\ 
& 5289, 5290, 12026, 10899, 16471, 16468, 17404, 17399, 17684, 15487 &\\
&16434, 15490, 15491, 16472,14274, 14275, 16469, 18432, 17407, 18437 &\\
&16466, 19877, 25677, 19968, 22829, 24609, 6159, 24600, 13108, 18436 &\\
&18430, 18427, 21839, 21830,  6163, 20642, 13417, 13413, 13412, 12027 &\\
&6160, 18428, 12029, 17406, 16438, 14273, 13111, 18439, 17401, 13415 &\\
&17410, 13113, 17409, 16437, 17402, 17403, 14272, 15492, 16467, 18429 &\\
& 19880, 16439, 16436, 16435, 17408, 18438, 17400, 13416, 15488, 15489 &\\
&13112, 13110, 16470, 13109, 13414,  5288, 20651,  3666,  5287,  5286 &\\
& 14270, 18435, 18431, 14271, 17411, 22838,17683,  6162,  6161, 25673 &\\
 & 26382, 25674, 26381, 16432, 15485, 14269, 18425, 17686, 17397, 17405 &\\
 &  18426, 18433, 14268, 16433, 18423, 17398, 13106, 17685, 15486, 13107&\\
 \hline
 Hydra-A &  575, 576, 2208 ,2330, 2331, 2332, 2333, 2334 ,4969 ,4970  & 297\,ksec \\ 
 \hline
 PKS0745-19 & 508, 510, 1383, 1509, 2427, 6103, 7694, 12881  & 272.9\,ksec \\
 \hline
\end{tabular}\label{table:1}
\end{table}

\begin{table}[b]
\caption{Summary of spatial and spectral binning for pGMCA.}
\centering
\begin{tabular}{c c c} 
 \hline\hline
 Cluster   & $\Delta$E  & Spatial binning\\ 
    & (keV)  & (arcseconds / pixels) \\ 
  (1)   & (2)  & (3) \\ 
 \hline
 Perseus & 0.0438 & 3.92 / 8~pixels\\ 
 M87  & 0.0438 &  1.96 / 4~pixels \\
 Centaurus & 0.2628 & 1.96 / 4~pixels\\ 
 Abell 2597 & 0.5  & 0.98 / 2~pixels \\ 
 Abell 1795 & 0.5  & 2.94 / 6~pixels \\ 
 Hydra-A &  0.1 &  1.96/ 4~pixels\\ 
 PKS0745-19 & 0.5 &  1.96 / 4~pixels \\
 \hline
\end{tabular}\label{table:summary-pGMCA}
\end{table}

\begin{table}
\caption{List of optical observations used in this paper}
\centering
\begin{tabular}{c c c c} 
 \hline\hline
Galaxy &  Instrument & OBSID & Exposure \\ \hline
 Perseus &  SITELLE &  15BE10 \& 17BC22 & 12210.9s \\ 
 Centaurus &  MUSE & 094.A-0859(A) \& 0103.A-0447(A) &  7271s \\ 
 M87 &  MUSE & 60.A-9312(A) &  5400s \\ 
  Abell\,2597 &  MUSE & 094.A-0859(A)  &  2700s \\
   Abell\,1795 &  MUSE & 094.A-0859(A)  &  2700s \\ 
 Hydra-A &  MUSE & 094.A-0859(A) &  2700s \\ 
 PKS0745-19 &  MUSE & 094.A-0859(A) &  2700s\\ 
 \hline
\end{tabular}
\label{table:MUSE}
\end{table}

\end{appendices}

\newpage\clearpage

\end{document}